# Design Strength-Ductility Synergy of Metastable High-Entropy Alloys by Tailoring Unstable Fault Energies

Xin Wang[1,†], Rafael Rodriguez De Vecchis[1,†], Chenyang Li[2,†], Soumya Sridar[1], Xiaobing Hu[3], Wei Chen[2,*], Wei Xiong[1,*]

*[1]Physical Metallurgy and Materials Design Laboratory, Department of Mechanical Engineering and Materials Science, University of Pittsburgh, Pittsburgh, PA, 15261, USA*
*[2]Department of Mechanical, Materials and Aerospace Engineering, Illinois Institute of Technology, Chicago, IL, 60616 USA*
*[3]Department of Materials Science and Engineering, Northwestern University Evanston, IL 60208, USA*

*[*] Corresponding Authors, Email: wchen66@iit.edu; w-xiong@outlook.com*
*[†] These authors contributed equally to this work*

## Abstract

Metastable alloys with transformation/twinning-induced plasticity (TRIP/TWIP) can overcome the strength-ductility trade-off in structural materials. Originated from the development of traditional alloys, the intrinsic stacking fault energy (ISFE) has been relied to tailor TRIP/TWIP in high-entropy alloys (HEA), but with limited quantitative success. Herein, we demonstrate a new strategy for designing metastable HEAs and validate its effectiveness by discovering seven new alloys with experimentally observed metastability for TRIP/TWIP. We propose unstable fault energies as the more effective design metric and attribute the deformation mechanism of metastable face-centered cubic alloys to UMFE (unstable martensite fault energy)/UTFE (unstable twin fault energy) rather than ISFE. Among the studied HEAs and steels, the traditional ISFE criterion fails in more than half of the cases, while the UMFE/UTFE criterion accurately predicts the deformation mechanisms in all cases. The UMFE/UTFE criterion provides a new paradigm for developing metastable alloys with TRIP/TWIP for enhanced strength-ductility synergy.

**Keywords:** high-entropy alloy, stacking fault energy, martensite, twinning, DFT

## Introduction

Developing alloys with a combination of high strength and ductility is the paramount goal in structural materials engineering. However, most strengthening mechanisms, such as precipitation and solid solution hardening, are detrimental to ductility[1,2]. Metastability-engineering[3–5] has been [6]demonstrated to be an effective strategy to overcome the strength-ductility trade-off in ferrous alloys[6] by introducing interface hardening from a dual-phase structure and transformation or twinning-induced plasticity (TRIP/TWIP). During deformation, martensite/twin formation provides alternative pathways for partial dislocations to glide, and the newly formed phase/twin boundary reduces the dislocation mean free path, leading to the dynamic Hall-Petch effect[7]. Recently, metastability-engineering has also been utilized in developing high-entropy alloys (HEAs)[4,5]that contain multiple principal elements[8–11]. The combination of 'HEA effects', such as the severe lattice distortion and solid solution strengthening, and the metastable engineering leads to HEAs with excellent mechanical properties[12]. Moreover, the concept of HEAs offers a vast composition space in which the microstructure and deformation mechanisms can be tuned to get superior properties. Nonetheless, the effective design of HEAs with desired microstructure and deformation mechanisms is still a formidable challenge.

HEAs represent a paradigm shift in materials research from the corner of phase diagrams to the central region of the high-dimensional phase space. Facing with the astronautical design space, computation-aided design offers a more efficient way than the Edisonian approach for exploring composition-process-structure relationships in HEAs[13–15]. For example, a variety of effective computational approaches, such as phenomenological parameters[16,17], machine learning (ML) models[15,18,19], and the CALPHAD (Calculation of Phase Diagrams) method[20–22], have been applied to predict the HEA phases for a given composition. However, the competition of phase stabilities is a function of both alloy composition and the processing history, especially for metastable phases. As a result, the influence of heat treatment and phase transformations, such as the athermal martensitic transformation during quenching, should be addressed to achieve higher prediction accuracy.

The competing deformation mechanisms in the face-centered cubic (fcc) phase, such as dislocation glide, twinning, and martensitic transformation, are usually determined by the intrinsic stacking fault energy (ISFE)[12,21]. The ISFE is the excess energy associated with the formation of the intrinsic stacking fault by the dissociation of a lattice dislocation $a_0/2$<110> into two $a_0/6$<112> partial dislocations[23,24] ($a_0$ is the lattice constant). In austenitic steels, TRIP prevails over TWIP when ISFE is lower than 20 mJ/m$^2$, and TWIP is found in alloys when the ISFE lies between 20 and 40 mJ/m$^2$ [25–27]. However, this rule does not generalize in other alloys such as HEAs. For example, transmission electron microscopy experiments found $Co_{10}Cr_{10}Fe_{40}Mn_{40}$ has a low ISFE of 13 ± 4 mJ/m$^2$, but this HEA is a TWIP alloy[28]. Moreover, recent studies have shown that the experiments tend to overestimate the ISFE in concentrated alloys[23], while the density functional theory (DFT) calculations can give negative ISFE values for TRIP HEAs[29,30], making the ISFE criterion impractical for rigorous alloy design and discovery. In fact, ISFE is the energy change after the stacking fault formation, but it does not necessarily correlate with the energy barrier for the martensite or twin formation process. A more in-depth understanding of the deformation-induced martensite/twin formation process and their energy barrier is thus critical. Previous experimental studies found that the formation of deformation twins is associated with the partial dislocation glide on every {111} plane in fcc materials, while hexagonal close-packed (hcp) martensite forms as the result of the movement of partial dislocations on every other {111} plane[31–33]. Therefore, we introduce two intrinsic quantities: unstable martensite fault energy (UMFE) and unstable twin fault energy (UTFE). The differences between UMFE/UTFE and ISFE define the energy barrier for martensite/twin formation, which controls the competition between different deformation mechanisms and allows predicting the TRIP/TWIP behaviors in HEAs.

Figure 1 illustrates our strategy for designing HEAs with the desired phase and deformation mechanism. First, we predict the phase stability using the CALPHAD method for more than 100,000 compositions of the Co-Cr-Fe-Mn-Ni system to screen single-phase fcc HEAs free of brittle intermetallics at the homogenization temperature of 1200˚C. Then, we compute the energy difference between the fcc and hcp phase at room temperature to identify the alloys as either fcc single phase or fcc + hcp dual phases. This energy difference describes the competition between these phases during quenching (Fig. 1b). Seven new alloys with TRIP/TWIP were selected based on the kernel density analysis (Fig. 1c) and their fcc stability compared with two reference alloys (Ref-1: $Co_{20}Cr_{20}Fe_{20}Mn_{20}Ni_{20}$, Ref-2: $Co_{10}Cr_{10}Fe_{40}Mn_{40}$). Then, we calculate the energy difference UMFE–ISFE and UTFE–ISFE to understand the energy barrier for martensite and twin formation, respectively (Fig. 1d). The comparison of the unstable fault energy difference defined in this work is considered as the new guidelines to introduce TRIP/TWIP effects. Lastly, we validate the design by conducting experiments and identifying the relationship between alloying elements, stacking fault energies, and deformation mechanisms (Fig. 1e). Herein, we demonstrate an effective design strategy for HEAs using metastability-engineering. This work reveals that the UMFE–ISFE and UTFE–ISFE predict the martensite and twin formation better than ISFE (Fig. 1f).

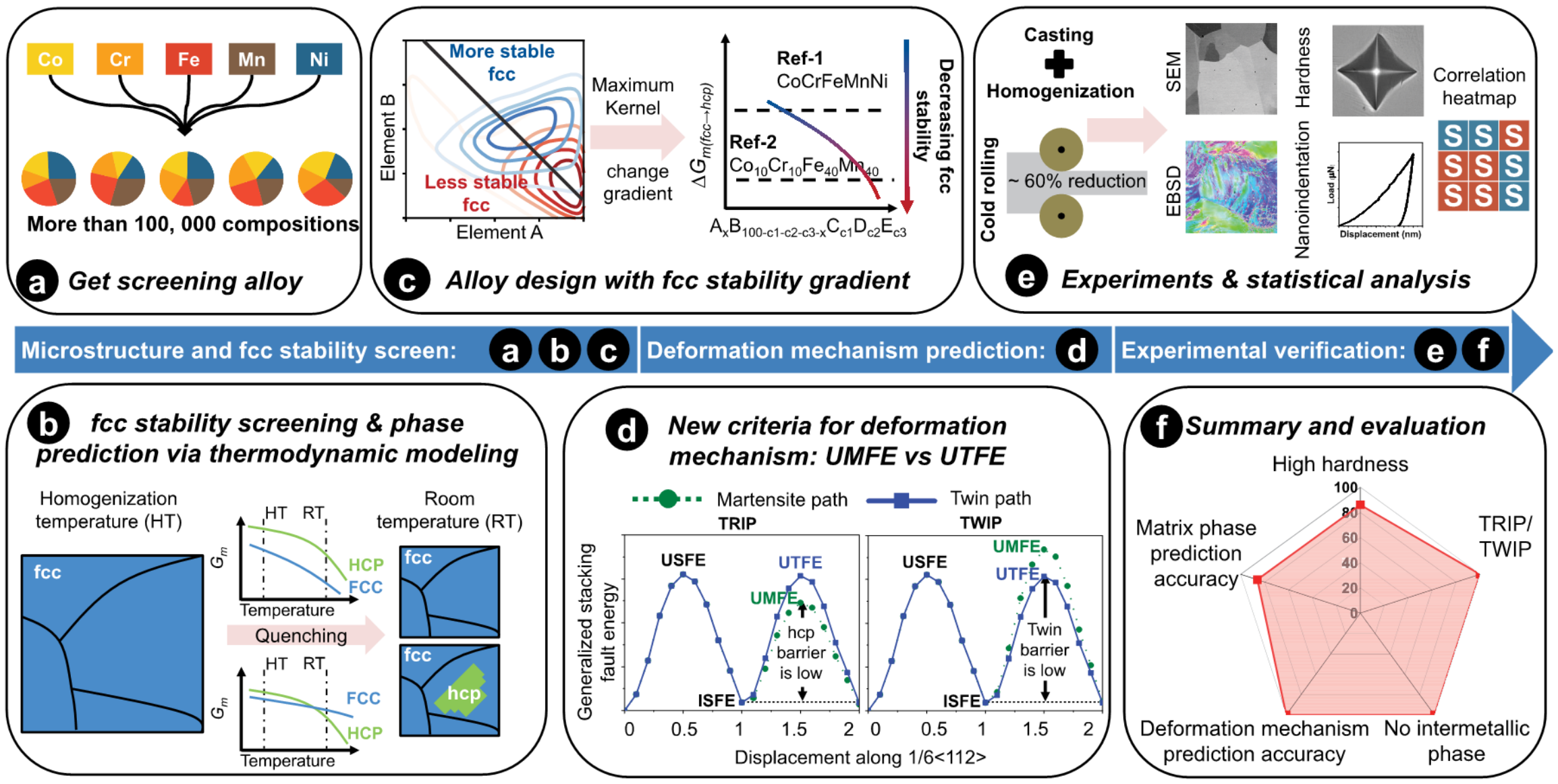

**Fig. 1 Design workflow. a-c** are schematics of the fcc stability and phase prediction via thermodynamic modeling. **a** Co, Cr, Fe, Mn, Ni are mixed into the HEA composition where all alloying elements are in the range of 0 to 40 at.%. **b** The phase at the homogenization temperature (1200˚C) and room temperature is predicted via thermodynamic modeling. We only consider alloys with a single fcc phase at 1200˚C for further calculations. If the Gibbs free energy of fcc is lower than hcp at the room temperature, the alloy is labeled as a fcc single-phase alloy. Otherwise, it is labeled as a fcc + hcp dual-phase alloy. **c** The Gibbs free energy difference between fcc and hcp for all compositions are compared with two reference alloys and are categorized into less stable fcc and more stable fcc. Based on the kernel density plot, the alloys with the compositionally complementary elemental pairs that produce the maximum change of kernel density are identified, and seven alloys with different fcc stability are chosen for further calculations. Since all designed alloys have lower fcc stability than Ref-1 ($Co_{20}Cr_{20}Fe_{20}Mn_{20}Ni_{20}$), they are expected to form martensite or twins after cold-rolling. **d** Deformation mechanism prediction *via* DFT. USFE, ISFE, UMFE, UTFE are unstable stacking fault energy, intrinsic stacking fault energy, unstable martensite fault energy, and unstable twin fault energy, respectively. We propose that if the energy barrier for martensite (hcp) formation (UMFE–ISFE) is smaller than the energy barrier for twining (UTFE–ISFE), the alloy is a TRIP alloy; Otherwise, the alloy will be TWIP dominant. **e-f** Design validation by experiments. **e** Schematics of sample preparation, testing, characterization, and statistical correlation analysis. **f** Evaluation of the design metrics. All the designed alloys have the secondary deformation mechanism and no intermetallic phase as predicted. More than 80 % of the designed alloys have higher hardness compared with reference HEA (Ref-2: $Co_{10}Cr_{10}Fe_{40}Mn_{40}$) and show the as-predicted microstructure.

# Methods

## *Calculation:*

**Thermodynamic screening.** The CALPHAD approach was used for thermodynamic phase screening. The calculations were performed using the TC-Python toolkit from Thermo-Calc software with the TCHEA4 database[34,35]. The searched composition space is 0 < Co ≤ 40 at.%; 0 < Cr ≤ 40 at.%; 0 < Mn ≤ 40 at.%; 0 < Ni ≤ 40 at.%, with Fe as the balance element and its upper limit was set to be 40 at.%.

**First-principles calculations**. DFT calculations with the exact muffin-tin orbital (EMTO)[36–39] method were performed to calculate the planar defect energies of the studied alloys. As an improved screened Koringa-Kohn-Rostoker (KKR) method, the EMTO method uses optimized overlapping muffin-tin potential spheres to represent the exact one-electron potential[40]. The one-electron equations were solved within the scalar relativistic approximation and the soft-core scheme. The chemical disorder of the HEA was treated with the coherent-potential approximation (CPA)[41–43]. The Perdew-Burke-Ernzerhof (PBE) generalized gradient approximation[44] was used to approximate the exchange-correlation energy. The EMTO Green's function was calculated self-consistently for 16 complex energy points distributed exponentially[45]. The stacking fault energy is calculated as the energy difference between the ideal fcc lattice and the modified lattice containing the stacking fault. To be specific, sliding one layer at different distances and fixing the relative positions of other layers will obtain a series of stacking fault energy. The sliding was modeled using tilted supercells from applying a partial dislocation burgers vector $n\mathbf{b}_p$ ($\mathbf{b}_p = a_0\langle 112\rangle/6$, where $a_0$ is the lattice constant, n increased by steps of 0.1 on the original supercell, which is parallel to the stacking direction (perpendicular to the stacking fault plane[49]. To eliminate the interaction between stacking fault layers and ensure convergence, a model of successive 9 (111) planes[45–48] was used in this work. For each 0.1 $\mathbf{b}_p$ sliding, the interlayer distances between the layers containing the stacking fault were optimized[49], while other interlayer distances were fixed. The *k*-mesh was carefully tested for convergence, and the $12 \times 24 \times 3$ mesh was adopted for all calculations.

**Correlation analysis.** All statistical analyses were coded using SciPy[50] and Scikit-learn[51]. For two continuous variables, the coefficient and associated *p*-value of Pearson correlation and Kendall rank correlation were calculated using SciPy[50]. For two binary variables pairs, the Phi coefficient was calculated using the Matthew correlation coefficient implemented in Scikit-learn[51]. Furthermore, the *p*-value was calculated using Fisher's exact test available in SciPy. For one binary variable and one continuous variable pair, the *p*-value was obtained from the Mann-Whitney rank test in SciPy, and the effect size was calculated based on the method proposed by Kerby[52].

### *Experiments:*

**Sample preparation and heat treatment.** High purity elements (Mn: 99.98 wt.%, Ni: 99.98 wt.%, Cr: 99.95 wt.%, Fe:99.99 wt.%, Co: 99.9 wt.%) were weighted in a high precision balance (± 0.1 mg), in accordance with the designed compositions indicated in Table 1. We added extra 2 wt.% of Mn to all compositions to compensate for the expected Mn evaporation losses during casting. 40 gram ingots for series one alloys and 10 gram ingots for the series two alloys were cast in an arc-melter (ABJ-338 manufactured, Materials Research Furnaces Inc., USA) operated with a copper crucible and under an argon atmosphere. Pure zirconium was used to remove any remnant oxygen present in the casting chamber. Moreover, each alloy was re-melted at least three times to ensure a homogenous ingot with less than 0.5% weight loss. Afterwards, samples disc-shapes and rectangular samples cut from the arc-melted ingots were encapsulated in quartz tubes, purged under vacuum, and backfilled with argon gas. Homogenization was carried out at 1200°C for one hour followed by ice water quenching. Disc-shaped samples were used to study the microstructure without deformation, while the rectangular samples were dedicated for cold rolling. 60% thickness reduction in a single pass during cold rolling was designed to be imposed in all alloys to ensure that large hcp plates or twins could form.

**Microstructure characterization.** Both undeformed and deformed samples were hot mounted, ground, and polished to a mirror-like finish using a 0.1 μm diamond and 0.02 μm silica particle suspension solutions. Microstructure characterization was conducted using an FEI Scios DualBeam scanning electron microscope (SEM). Energy dispersive spectroscopy (EDS) was carried out to ensure no element segregation was present after homogenization. Furthermore, electron backscatter diffraction (EBSD) was performed for a 400 μm × 400 μm area with a step size of 0.75 μm for both sets of samples and analyzed with the TSL OIM Analysis™ v8 software.

**Microhardness and nanoindentation.** Vickers microhardness measurements were conducted in all samples using an AMH55 with LM310AT Microindenter, LECO Corporation automatic hardness tester with a 100 gram-force and 10 second dwell time. Reported values correspond to an average and standard deviation of ten readings. Nanoindentation was carried out in the polished homogenized samples using a Hysitron TI 950 TriboIndenter with a standard Berkovich diamond probe. A triangular load function with a peak load of 2000 μN and constant 200 μN/s loading/unloading rate was implemented to perform a grid of 10 x 3 indents, spaced 15 μm from one another to avoid overlap of indents. The reported hardness and reduced modulus correspond to the average of thirty readings.

## Results and Discussion

**Microstructure and fcc stability screening using thermodynamic modeling.** To perform a high-fidelity prediction on phase stability, we evaluated the reliability of the CALPHAD modeling for HEAs by comparing between experiments and model predictions. Figure 2a summarizes the accuracy of reported phase predictions using machine learning[15,18,19], CALPHAD[18,22], and phenomenological parameters[18] (further details can be found in Supplementary Table 1). It is clear that the machine learning and CALPHAD approaches perform better than the phenomenological parameters[18]. Both machine learning and CALPHAD approaches are not computationally intense methods. Although machine learning models reported in the literature show satisfactory accuracy, heat-treatment conditions are often not considered as model inputs. Existing machine learning classification models were also not built to predict specific phases and phase fractions[15,19], that are necessary for a finer alloy design task. On the contrary, the CALPHAD approach calculates the phase and their fraction for a given temperature and composition, and the phase accuracy can be further improved as the previous study did not consider the athermal phase transformation during cooling[18,22].

Since the primary product of martensitic transformation in fcc HEAs is the ε martensite with a hcp structure[4,53,54], the chemical driving force of martensitic transformation can be expressed as the Gibbs energy change ($\Delta G_{m(fcc \rightarrow hcp)}$) that corresponds to the thermodynamic model predicted ISFE[55]. Smaller $\Delta G_{m(fcc \rightarrow hcp)}$ and ISFE indicate lower fcc stability, i.e., it is easier for fcc to transform into hcp. We collected experimentally measured stacking fault energies, which usually have an uncertainty of around 5 $mJ/m^2$ (details in Supplementary Table 2), and compared them with the calculated $\Delta G_{m(fcc \rightarrow hcp)}$ at room temperature (293 K) (Fig. 2b). Based on the Pearson correlation coefficient $r$, the experimental ISFE and $\Delta G_{m(fcc \rightarrow hcp)}$ show a very weak linear relationship for all collected HEAs since $r$ is close to 0. Meanwhile, the $r$ for alloys containing Co, Cr, Fe, Mn, and Ni is close to 1, and the $p$-value is smaller than 0.05. This result shows a statistically significant positive correlation between the experimental ISFE and the CALPHAD calculated $\Delta G_{m(fcc \rightarrow hcp)}$. Thus, the CALPHAD approach is expected to be reliable for predicting the fcc to hcp phase transformation in the Co-Cr-Fe-Mn-Ni systems.

Next, we performed a high-throughput screening of more than 100,000 compositions in the Co-Cr-Fe-Mn-Ni system. Compositions with a single fcc phase at the homogenization temperature (1200˚C) were selected for further study. This screening criterion ensures that the phase composition is the same as the overall alloy composition and no ductility degradation is caused by the formation of brittle intermetallic phases. Figures 2c & d are violin plots that show the elemental distribution of fcc single-phase ($\Delta G_{m(fcc \rightarrow hcp)}$ at 293 K > 0) and fcc + hcp dual-phase ($\Delta G_{m(fcc \rightarrow hcp)}$ at 293 K < 0) HEAs, respectively. We find that fcc single-phase HEAs have less dependency on any particular alloying element. However, the fcc + hcp dual-phase HEA is more likely to form with a high Co or Cr content and a limited addition of Mn or Ni.

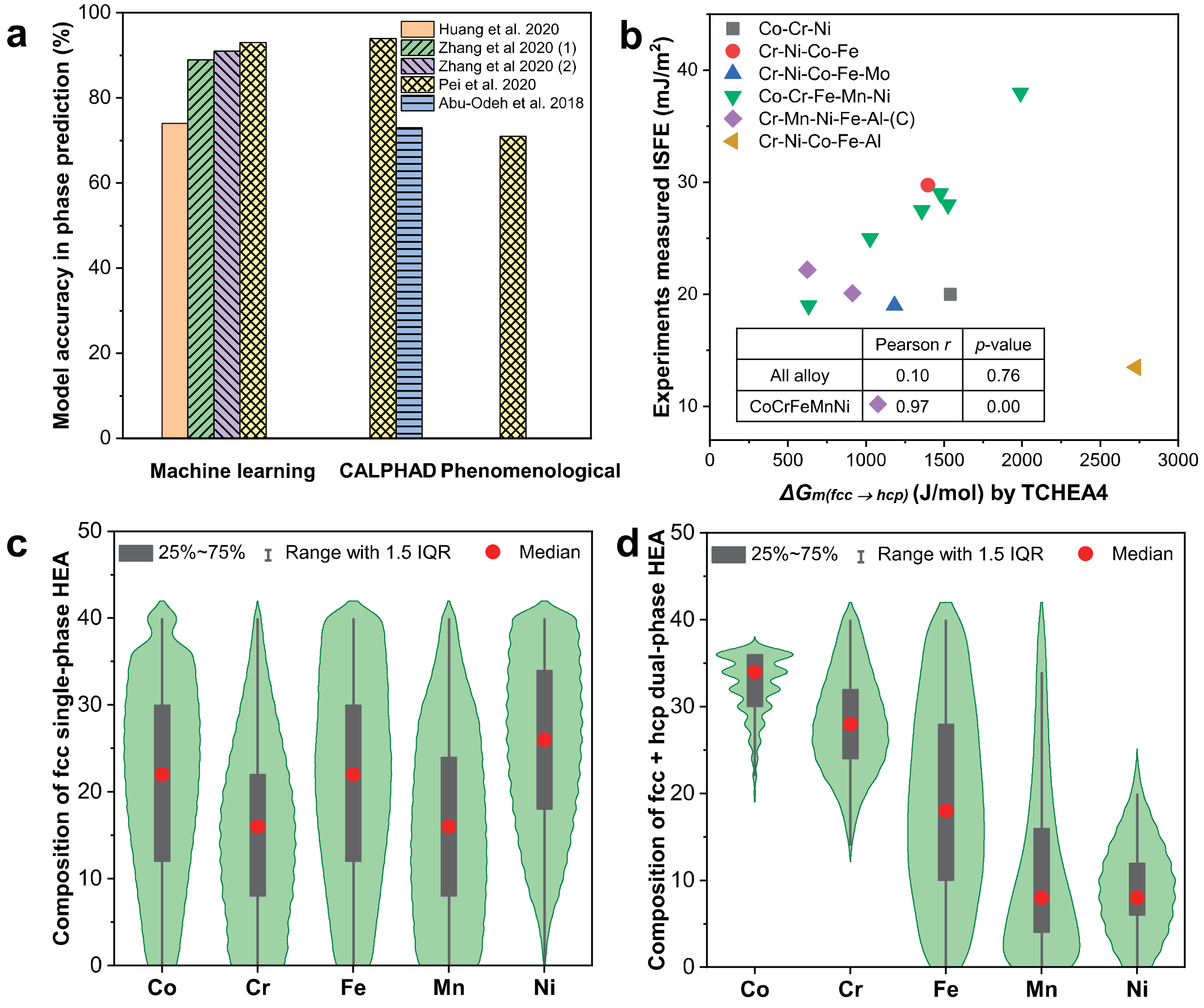

**Fig. 2 Evaluation of CALPHAD approach and thermodynamic model screening results. a** Comparison of the phase prediction accuracy using machine learning, CALPHAD, and phenomenological parameter approach based on literature, where the classification type, methods, and accuracy are available in Supplementary Table 1[15,18,19,22]. **b** Comparison of the experimentally measured intrinsic stacking fault energy (ISFE)[56–63] and the model-predicted value of the Gibbs free energy change of fcc ⟶ hcp transformation at room temperature. The inset table shows the Pearson correlation coefficient *r* for all data, and alloys containing Co, Cr, Fe, Mn, Ni. **c-d** Screening results of $Co_xCr_yFe_{(100-x-y-z-w)}Mn_zNi_w$ (0 ≤ x,y,z,w, 100-x-y-z-w ≤ 40 atomic fraction), where the alloy compositions form single fcc at 1200˚C. The width of the green area represents the probability density of compositions corresponding to different alloying elements, and 1.5IQR is 1.5 times the interquartile range. **c** The distribution of fcc single-phase alloy, which has $\Delta G_{m(fcc\rightarrow hcp)} > 0$ at 293 K, **d** The distribution of fcc + hcp dual-phase alloy, which has $\Delta G_{m(fcc\rightarrow hcp)} < 0$ at 293 K. The calculations of a few hundred compositions failed due to the numerical issue and they were not included in the data visualization.

It is widely accepted that lower ISFE, *i.e.*, the decreasing fcc stability, leads to the deformation mechanisms switching from a pure dislocation glide to a combination of a dislocation glide and twinning followed by a dislocation glide together with martensitic transformation[23,64]. Hence, identifying alloys with gradient fcc stability could help design alloys with TRIP and TWIP. Two well-studied HEAs were used as the reference to guide our design. The first reference alloy is $Co_{20}Cr_{20}Fe_{20}Mn_{20}Ni_{20}$ HEA, which has a more stable fcc that only shows deformation-induced twin after severe[65] or low-temperature deformation[11,66]. The second reference alloy is $Co_{10}Cr_{10}Fe_{40}Mn_{40}$ with a less stable fcc, which shows TWIP during room temperature deformation[28,67] and hcp martensite formation after high-pressure torsion[68] or low-temperature deformation[69]. By comparing the calculated fcc stability ($\Delta G_{m(fcc\rightarrow hcp)}$) of the screened alloys with $Co_{20}Cr_{20}Fe_{20}Mn_{20}Ni_{20}$ ($\Delta G_{m(fcc\rightarrow hcp)}^{\mathrm{Co20Cr20Fe20Mn20Ni20}}$) and $Co_{10}Cr_{10}Fe_{40}Mn_{40}$ ($\Delta G_{m(fcc\rightarrow hcp)}^{Co10Cr10Fe40Mn40}$), the screened alloys are classified as alloys with less stable fcc and more stable fcc. The kernel density estimation was generated based on the number of labeled alloys with the Gaussian kernel, and it was applied to determine the alloy probability density distribution of two varying elements. In total, ten kernel density plots were generated. The plots of Co-Cr, Fe-Ni, Fe-Cr, and Co-Ni are shown in Fig. 3a-d, and the rest of plots are shown in Supplementary Fig. 1. In the Fe-Cr and Co-Ni pairs, we can shift the alloys from more stable fcc to less stable fcc by increasing the composition of one element and reducing the other element content as illustrated by the red and black lines in Figs. 3c and 3d, respectively. As a result, we selected seven samples from the Co-Ni (series 1) and Fe-Cr (series 2) series that have a monotonic reduction in $\Delta G_{m(fcc\rightarrow hcp)}$ for further DFT and experimental study (Fig. 3e). The calculated and experimental results of the seven designed alloys and two reference alloys are summarized in Table 1.

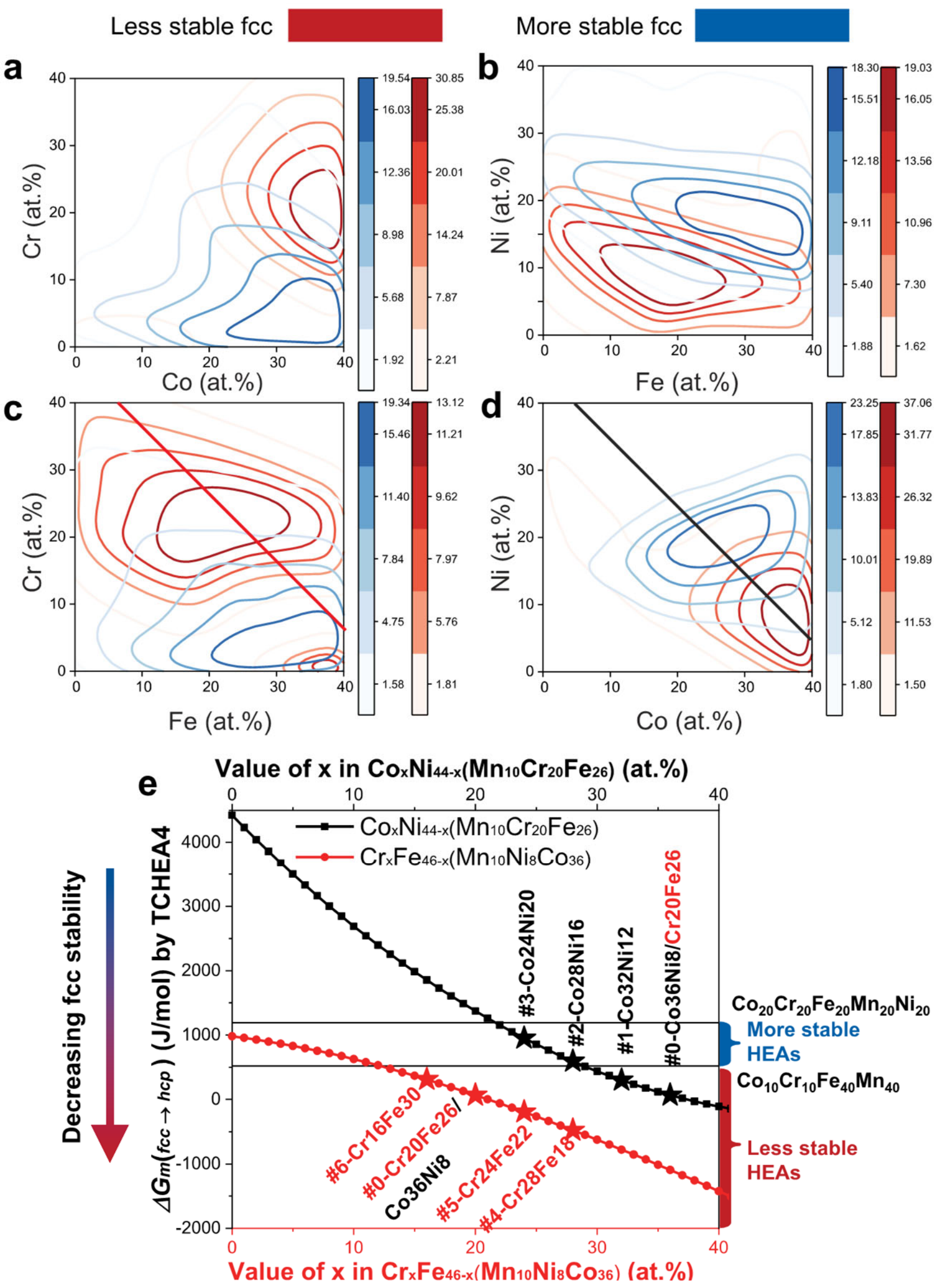

**Fig. 3 Design of TRIP/TWIP HEAs using kernel density plot with different degrees of fcc instability.** The screened alloys have been classified as less stable fcc alloy with red color ($\Delta G^{HEAs}_{m(fcc\rightarrow hcp)} < \Delta G^{Co10Cr10Fe40Mn40}_{m(fcc\rightarrow hcp)}$), and more stable fcc alloy with blue color ($\Delta G^{Co10Cr10Fe40Mn40}_{m(fcc\rightarrow hcp)} < \Delta G^{HEAs}_{m(fcc\rightarrow hcp)} < \Delta G^{CoCrFeMnNi}_{m(fcc\rightarrow hcp)}$). The kernel density distribution of the two classes of alloys has been plotted with two varying elements. A darker contour line indicates higher kernel density of the alloys in the composition region. **a** Co-Cr pair and **b** Fe-Ni pair where the increase of one element and decrease of another element do not lead to the kernel density change from more stable fcc to less stable fcc. **c** Fe-Cr pair and **d** Co-Ni pair where the increase of one element and decrease of another element lead to the kernel density change from more stable fcc to less stable fcc. The straight lines indicate the alloy design path. **e** the composition and $\Delta G_{m(fcc\rightarrow hcp)}$ of the designed alloys and the reference alloys. The red and black curves correspond to the straight lines in c and d of the same color. The alloys with star marks are selected for further study. Results of other pairs such as Mn-Ni, Fe-Mn, Co-Mn, Fe-Co, Cr-Mn, and Cr-Ni are available in Supplementary Fig. 1.

**Table 1 Summary of selected alloy compositions in at.% (atomic percentage), calculated properties, and experimental observations.**

| Label | Alloy composition (at.%) | Stable phases at 1200˚C | $\Delta G_{m(fcc \rightarrow hcp)}$ at room temperature (J/mol) | Prediction | | Experiments | |
|---|---|---|---|---|---|---|---|
| | | | | Phase at room temperature (CALPHAD) | Dominant secondary Deformation mechanism (DFT) | Phase at room temperature | Dominant secondary Deformation mechanism |
| *Reference alloys from literature* | | | | | | | |
| Ref-1 | $Co_{20}Cr_{20}Fe_{20}Mn_{20}Ni_{20}$ | fcc | 1193 | fcc | TWIP | fcc | TWIP (Tensile, 77 K[11,66]; Cold draw, 293 K[65]) |
| Ref-2 | $Co_{10}Cr_{10}Fe_{40}Mn_{40}$ | fcc or fcc + sigma | 519 | fcc or fcc + sigma | TWIP | fcc | TWIP (Tensile, 293K[28,67]) |
| *Series 1 Increasing Ni, decreasing Co* | | | | | | | |
| #0-Co36Ni8/ Cr20Fe26 | $Co_{36}Cr_{20}Fe_{26}Mn_{10}Ni_{8}$ | fcc | 67 | fcc | TRIP | fcc | TRIP |
| #1-Co32Ni12 | $Co_{32}Cr_{20}Fe_{26}Mn_{10}Ni_{12}$ | fcc | 300 | fcc | TWIP | fcc | TWIP |
| #2-Co28Ni16 | $Co_{28}Cr_{20}Fe_{26}Mn_{10}Ni_{16}$ | fcc | 593 | fcc | TWIP | fcc | TWIP |
| #3-Co24Ni20 | $Co_{24}Cr_{20}Fe_{26}Mn_{10}Ni_{20}$ | fcc | 952 | fcc | TWIP | fcc | TWIP |
| *Series 2 Increasing Cr, decreasing Fe* | | | | | | | |
| #4-Cr28Fe18 | $Co_{36}Cr_{28}Fe_{18}Mn_{10}Ni_{8}$ | fcc | -478 | fcc + hcp | TRIP | fcc + hcp | TRIP |
| #5-Cr24Fe22 | $Co_{36}Cr_{24}Fe_{22}Mn_{10}Ni_{8}$ | fcc | -196 | fcc + hcp | TRIP | fcc + hcp | TRIP |
| #0-Cr20Fe26/ Co36Ni8 | $Co_{36}Cr_{20}Fe_{26}Mn_{10}Ni_{8}$ | fcc | 67 | fcc | TRIP | fcc | TRIP |
| #6-Cr16Fe30 | $Co_{36}Cr_{16}Fe_{30}Mn_{10}Ni_{8}$ | fcc | 310 | fcc | TRIP | fcc + hcp | TRIP |

**Prediction of deformation mechanism from first-principles calculations**. Figure 4a-g illustrates the initial fcc lattice and different stacking fault structures formed during the passage of $a_0$/6<112> Shockley partial dislocations. Our model is a 9-layer supercell where the ISF forms at the boundary of neighboring supercells. After 0.5$\mathbf{b}_p$ ($\mathbf{b}_p$ is the burgers vector of Shockley partial dislocation) shear displacement of the upper layers above layer 9, the generalized stacking fault energy reaches the first local maxima that is termed by Rice as the unstable stacking fault energy (USFE) [70], which can be treated as the energy barrier of forming an ISF[71]. After one $\mathbf{b}_p$ displacement, the energy reaches the local minima (ISFE) that can also be determined by experiments such as the TEM or XRD. There are two possible further shearing pathways. One is shearing with successive (111) planes. The continuing passage of partial dislocation starts from the 11th plane and will reach a local energy maximum with 1.5$\mathbf{b}_p$ called the UTFE. The energy difference between UTFE and ISFE is the energy barrier for twin nucleation. The second local minimum energy corresponds to the extrinsic stacking fault (ESF), and it can be viewed as a twin embryo (TE). Repeating the above process in the following layers will result in a larger twin, and the UTFE maintains a similar value[47]. Unlike twin formation, the mechanism of hcp martensite formation is not well understood. Based on reported experimental observations[31–33], we propose that another possible shearing path is partial dislocation shearing on every other (111) plane, where the passage of the partial dislocation starts from the $12^{th}$ layer. As a result, the local minimum energy structure is a martensite embryo (ME), and the energy peak at 1.5$\mathbf{b}_p$ is defined as the UMFE. The energy difference between UMFE and ISFE is the energy barrier for martensite embryo formation. The energies corresponding to the atomic configuration shown in Fig.4a-g for the two reference alloys and the seven designed alloys are presented in Fig. 4h. We find that the USFE and UTFE are similar for all studied alloys, while the UMFE are different from UTFE. Since the deformation mechanism is favorable when the energy barrier is low, we hypothesize that martensitic phase transformation prevails when UMFE is smaller than UTFE, and deformation twinning dominates if the UMFE is larger than UTFE. Such predictions for the designed alloy and two reference alloys are listed in Table 1.

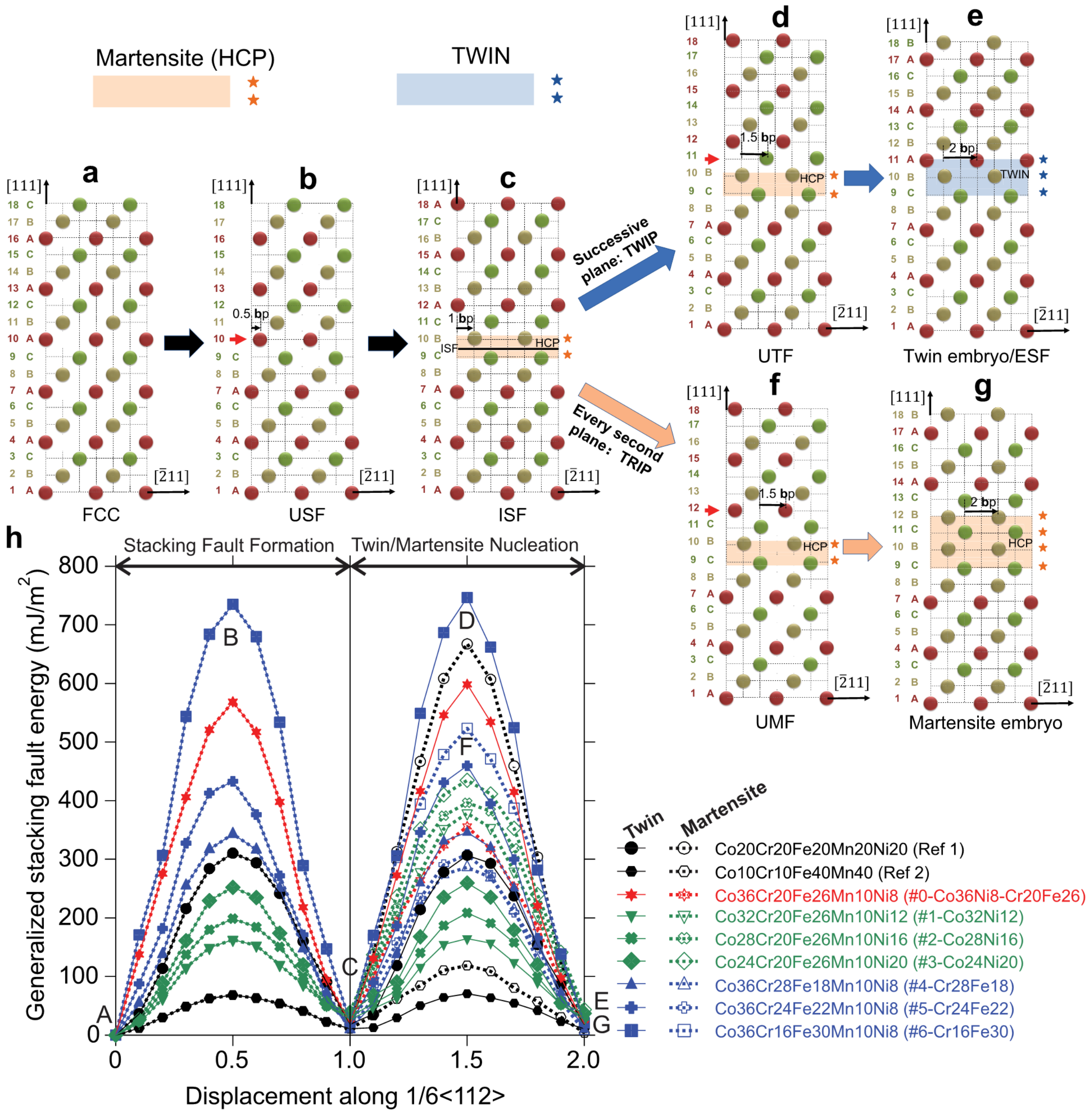

**Fig. 4 Calculation of stacking fault energies for martensite and twin formation and the corresponding atomic configuration.** Atomic configuration of **a** fcc structure with the ABCABC stacking sequence, **b** unstable stacking fault (USF), **c** intrinsic stacking fault (ISF), **d** unstable twin fault (UTF), **e** twin embryo (TE) or extrinsic stacking fault (ESF), **f** unstable martensite fault (UMF), and **g** martensite embryo (ME). Red, brown, and green dots represent atoms in fcc alloys corresponding to the A, B, and C layers, respectively. The orange square and stars represent the atomic layers that can be viewed as the hcp phase. The blue square and stars represent a twin structure. All atomic structures are formed through the shear displacement with a Shockley partial dislocation $\mathbf{b}_p = a_0/6<112>$. **h** Calculated stacking fault energies of two reference alloys and seven designed alloys using DFT. The local minima and maxima noted with uppercase A, B, C, etc., are the energies that correspond to the structure shown in subfigure a, b, c, etc., respectively.

**Experimental verification and TRIP/TWIP mechanism analysis.** The seven designed alloys were prepared by arc melting and homogenized at 1200˚C, followed by water quenching. We measured the composition of the homogenized samples using energy-dispersive spectroscopy (EDS), and it is found that the sample compositions are close to the designed compositions (Supplementary Tables 3-9). Moreover, the EDS mapping reveals that the segregation in as-cast samples was removed after heat treatment, leading to a uniform distribution of alloying elements in the homogenized sample (Supplementary Figs. 4-10).

Figure 5 summarizes the microstructure of the homogenized and cold-rolled samples characterized by electron backscatter diffraction (EBSD). The phase fractions before and after deformation are listed in Supplementary Table 10. Based on Fig. 5a1-5g1, the $Co_xCr_{44-x}Fe_{26}Mn_{10}Ni_8$ (x = 36, 32, 28, 24) alloys are fcc single-phase, and the $Co_{36}Cr_xFe_{46-x}Mn_{10}Ni_8$ (x = 28, 24,16) alloys are fcc + hcp dual-phase HEAs. From Table 1, it is evident that the predicted and observed phases are in good agreement for six out of seven designed alloys, indicating that the phase screening strategy in this work is reliable. By comparing the phase map of homogenized (Fig. 5a1-5g1) and deformed samples (Fig. 5a3-5g3), we find that all designed alloys show the existence of deformation-induced hcp martensite and minor body-centered cubic (bcc) martensite. Moreover, the $Co_{36}Cr_xFe_{46-x}Mn_{10}Ni_8$ (x = 28, 24, 20, 16) samples show a noticeable increase in the hcp phase fraction after deformation without any twin formation. This phenomenon implies that the martensitic transformation is the dominant deformation mechanism besides dislocation glide in these alloys. On the other hand, the $Co_xCr_{44-x}Fe_{26}Mn_{10}Ni_8$ (x = 32, 28, 24) HEAs show the Σ3 twin-boundaries in grain boundary maps (red lines in Figs. 5b5, 5c5, 5d5), and limited hcp formation when compared with TRIP dominant alloys. This result suggests that twining dominates during the deformation of these alloys, and we have labeled them as the TWIP dominant alloys. Figure 5a6-5g6 are the kernel average misorientation (KAM) maps for the deformed samples. The grain, twin, and phase boundaries show higher KAM values, indicating a higher density of geometrically necessary dislocations (GND)[72]. By comparing these phase maps (Fig. 5a3-5g3), it is noticed that the bcc phase usually forms at the intersection between hcp phases with a very high KAM value. Such a microstructure confirms that the hcp phase is the precursor for bcc martensite formation, which is in agreement with the previous study[73]. The deformation mechanisms confirmed with EBSD are also summarized in Table 1, and they agree with the DFT prediction.

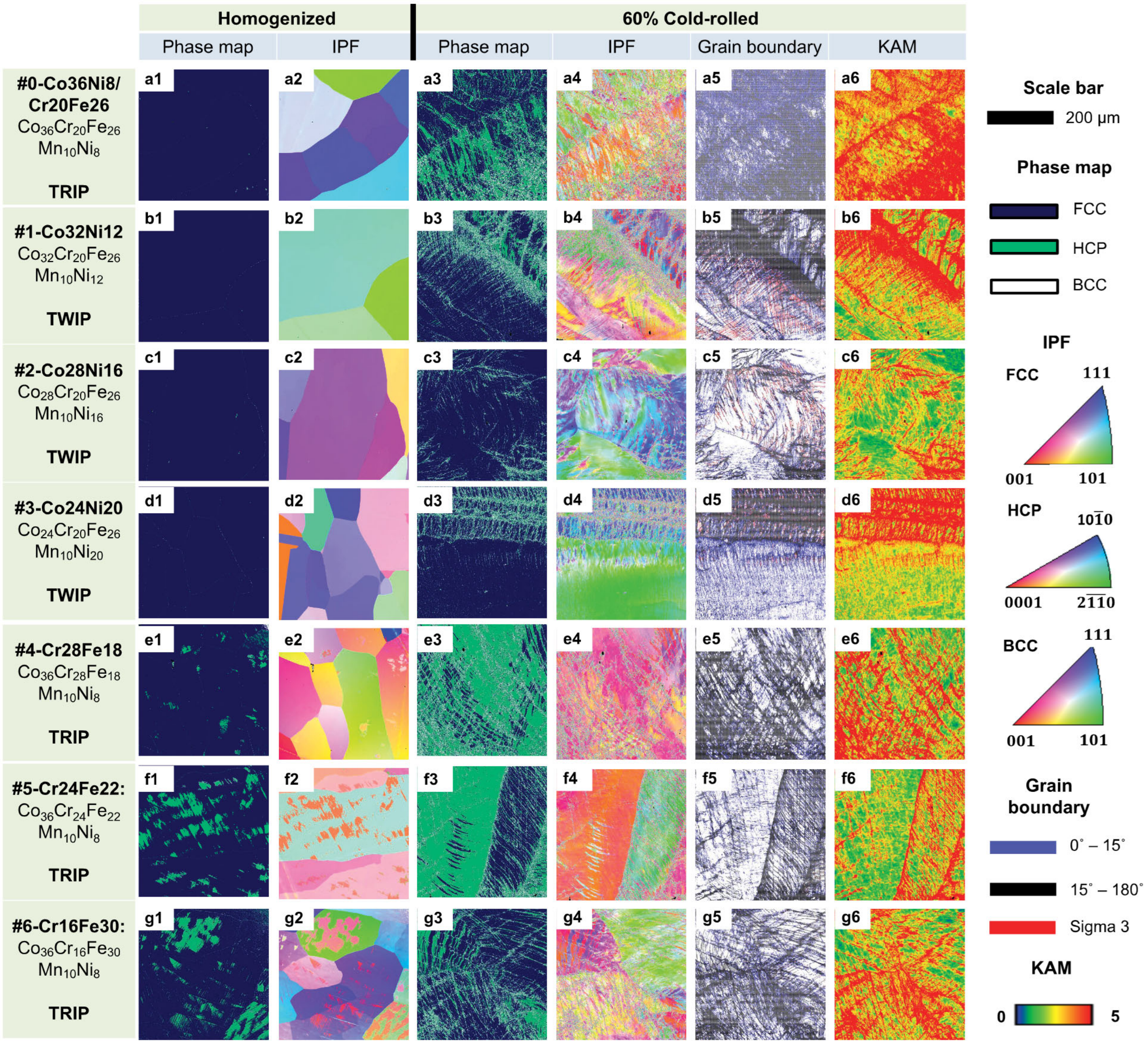

**Fig. 5 EBSD characterization of designed alloys.** Microstructure of **a** $Co_{36}Cr_{20}Fe_{26}Mn_{10}Ni_8$, **b** $Co_{32}Cr_{20}Fe_{26}Mn_{10}Ni_{12}$, **c** $Co_{28}Cr_{20}Fe_{26}Mn_{10}Ni_{16}$, **d** $Co_{24}Cr_{20}Fe_{26}Mn_{10}Ni_{20}$, **e** $Co_{36}Cr_{28}Fe_{18}Mn_{10}Ni_8$, **f** $Co_{36}Cr_{24}Fe_{22}Mn_{10}Ni_8$, **g** $Co_{36}Cr_{16}Fe_{30}Mn_{10}Ni_8$. Column **1** is phase map and column **2** is Inverse pole figure (IPF) map of the homogenized samples. Column **3** is phase map, column **4** is IPF map, column **5** grain is boundary map, and column **6** is Kernel Average Misorientation (KAM) map of the samples after cold rolling. The legends for all figures and the scale bar are listed on the right side of the figure.

In order to understand the correlation between alloying elements, deformation mechanisms, stacking fault energies, and the previously proposed deformation mechanism criteria[74], three correlation coefficients were calculated based on the variable type and visualized as a heatmap (Fig. 6). The correlation coefficient ranges from -1 to 1, where 1 represents a perfect positive correlation, 0 corresponds to no monotonic correlation, and -1 gives a perfect negative correlation between the parameter pair. We focus on the statistically significant results with a $p$-value smaller than 0.05. Based on the heatmap, we can study the correlation between composition and the stacking fault energies. Co has a strong positive correlation with USFE and UTFE, which implies Co increases the energy barrier for stacking fault and twin formation. Furthermore, Co decreases ISFE, ESFE/TEE, UMFE, MEE, although the effect is not statistically significant. This observation is consistent with the previous study suggesting that adding Co can decrease ISFE and shift the deformation mechanism from twining to martensitic transformation[75]. However, it is noted that higher Co content leads to a higher USFE, which may limit the formation of stacking fault that serves as the precursor for the martensite and twin formation. Another important element is Ni, which exhibits a strong effect on increasing the ISFE, ESFE/TEE, UMFE, and the MEE. Therefore, while Ni has a minor impact on the twinning energy barrier, its content must be limited to promote the TRIP effect. The findings of the effect of Ni on different stacking fault energies offer a fundamental explanation to the literature reported phenomenon[76] that adding Ni changes the deformation mechanism from martensitic transformation to twinning. As we have proposed, the competition between twinning and martensitic transformation is governed by the energy barriers, such as UTFE–ISFE and UMFE–ISFE. These criteria can be simplified by comparing the value of UTFE and UMFE to predict the dominant deformation mechanism besides the dislocation glide. The condition UMFE $<$ UTFE has a strong positive correlation with TRIP, and a strong negative correlation with TWIP. These results suggest that the UMFE $<$ UTFE criterion can predict the deformation mechanism accurately. On the contrary, the parameters typically used in previous studies, such as ISFE[25–27] and twinnability[54,74,77], do not show a statistically significant correlation with the deformation mechanisms.

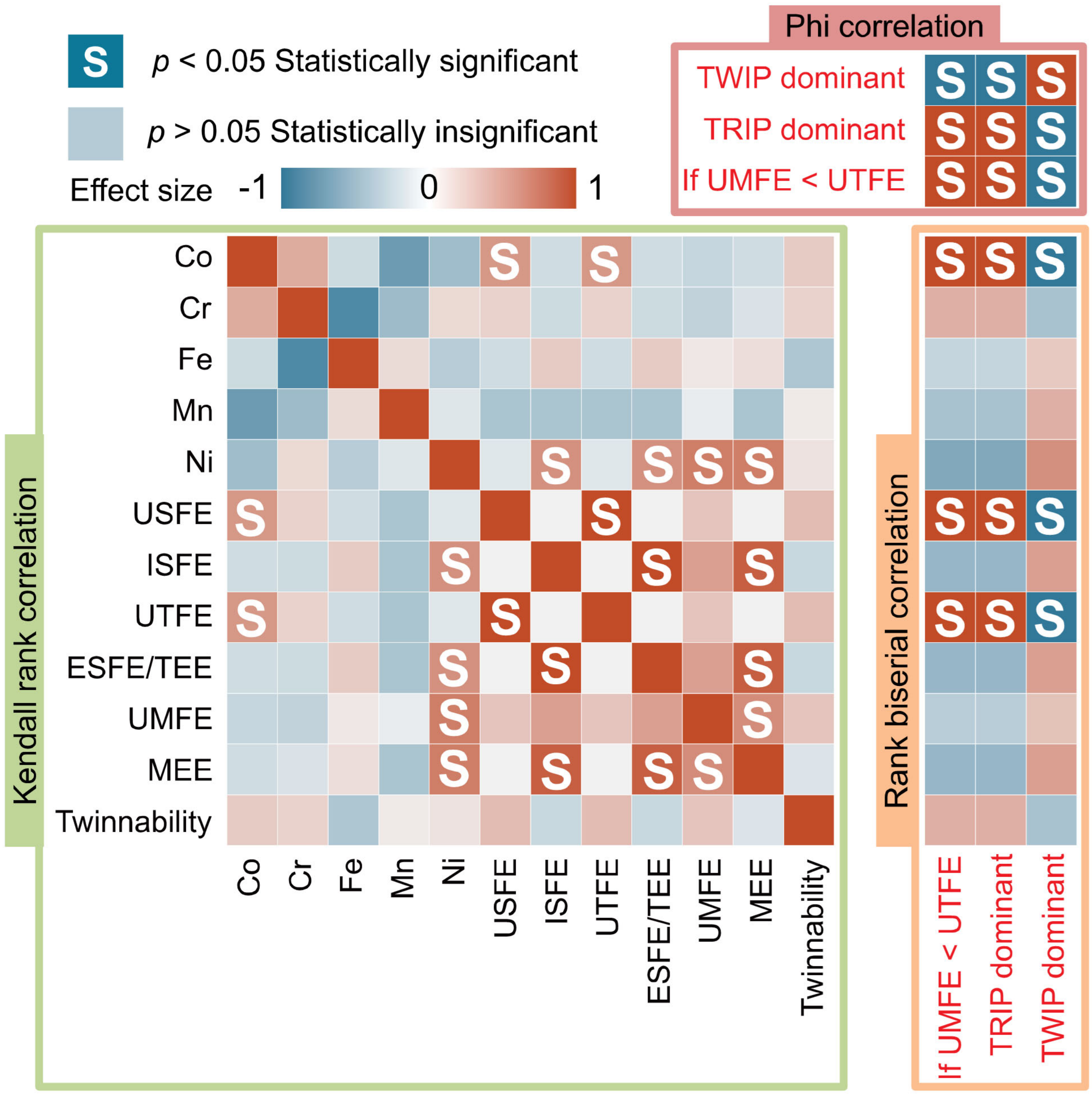

**Fig. 6 Correlation coefficient heat map of features and their relationship with TRIP and TWIP.** Features in red font are the binary variables, while features in black font are continuous variables. The Phi coefficient (red box), Kendall rank correlation coefficient (green box), and rank biserial correlation coefficient (orange box), and *p*-value are calculated for binary-binary, continuous-continuous, and binary-continuous parameter pairs. The red/blue color bar describes the correlation effect size, which indicates a positive/negative relationship between two features, and the darker color means the stronger relationship. Statistically significant results with a *p*-value less than 0.05 are marked with white S. The correlation significances between elements are not included for simplicity. The features can be divided into several categories: Composition (Co, Cr, Fe, Mn, Ni), DFT calculated features (USFE, ISFE, UTFE, ESFE, UMFE, MNE), literature reported criteria for determining deformation mechanism (Twinnability[54,74,77]), criteria for determining TRIP/TWIP proposed in this work (If UMFE<UTFE), and the experimentally observed deformation mechanism (TRIP dominant and TWIP dominant).

To better understand the deformation mechanisms and stacking fault energies, we compared the local minimum energies, ISFE, ESFE, and MEE in Fig 7a1, and the energy barriers, USFE, UTFE–ISFE, and UMFE–ISFE, in Fig. 7b1. We also plotted the hardness of the homogenized and deformed samples along with the hardness values reported in literature for Ref-1 ($Co_{20}Cr_{20}Fe_{20}Mn_{20}Ni_{20}$)[65,78–83] and Ref-2 ($Co_{10}Cr_{10}Fe_{40}Mn_{40}$)[84] alloys (Fig. 7c1). All local minimum energies are positive and smaller than 50 mJ/m$^2$, which is much lower than the energy barriers. Alloys with an increased Cr and Co content show a lower ISFE, which agrees with the thermodynamic model prediction of fcc instability in Fig. 3e. However, while the CALPHAD-based thermodynamic model shows that all the designed alloys have lower fcc stability than the Ref-1 alloy, the DFT calculations show that some designed alloys have higher ISFE, implying these alloys have a higher fcc stability than the Ref-1 alloy. Such discrepancies can be reduced by introducing a more advanced thermodynamic model considering interfacial energy and improving thermodynamic databases[27,85]. The alloys noted as TRIP dominant have lower martensite embryo energy (MEE) than the ISFE and ESFE/TEE, indicating the martensite formation introduces less energy into the system than the formation of stacking faults or twins. The TWIP dominant alloys also have a lower ESFE/TEE, suggesting a relatively stable twin structure, except for the Ref-2 alloy. Moreover, we have plotted two dashed horizontal lines at 20 and 40 mJ/m$^2$ that are widely adopted as the ISFE upper limits for TRIP and TWIP[25–27], respectively. However, the TRIP dominant $Co_{36}Cr_xFe_{46-x}Mn_{10}Ni_8$ (x = 16, 20) alloys have an ISFE higher than 20 mJ/m$^2$, and the TWIP dominant Ref-1 and Ref-2 alloys exhibit a much lower ISFE. Instead, the martensitic transformation energy barrier (UMFE–ISFE) for TRIP dominant alloys is much smaller than the stacking fault and the twin formation energy barriers (USFE and UTFE–ISFE). We found that the TRIP dominant HEAs have higher hardness compared to the TWIP dominant HEAs. Moreover, the hardness increased after deformation, showing a high strain hardening effect, and the hardness of the cold-rolled sample tends to decrease with the higher energy barriers for martensite/twin formation.

The results are summarized in the contingency tables embedded in Figs. 7a2 and 7b2 to elucidate the correlation between fault energies and deformation mechanisms. All samples have a low ISFE that is lower than 40 mJ/m$^2$. This implies that a low ISFE, *i.e.*, low fcc stability, is favorable for introducing secondary deformation mechanisms, such as martensitic transformation and twinning. However, we find that only two of the four TRIP dominant alloys show an ISFE smaller than 20 mJ/m$^{2,}$ and three of the five TWIP alloy have an ISFE that lies within the range of 20-40 mJ/m$^2$, which means the accuracy of this criteria is only about 56 %. On the contrary, the accuracy of our model for predicting the deformation mechanism, based on the UMFE–ISFE and UTFE–ISFE relationships, is 100%. Our finding proves that the UMFE–ISFE and UTFE–ISFE, rather than ISFE, should be used as a more general criterion to determine the deformation mechanisms in HEAs.

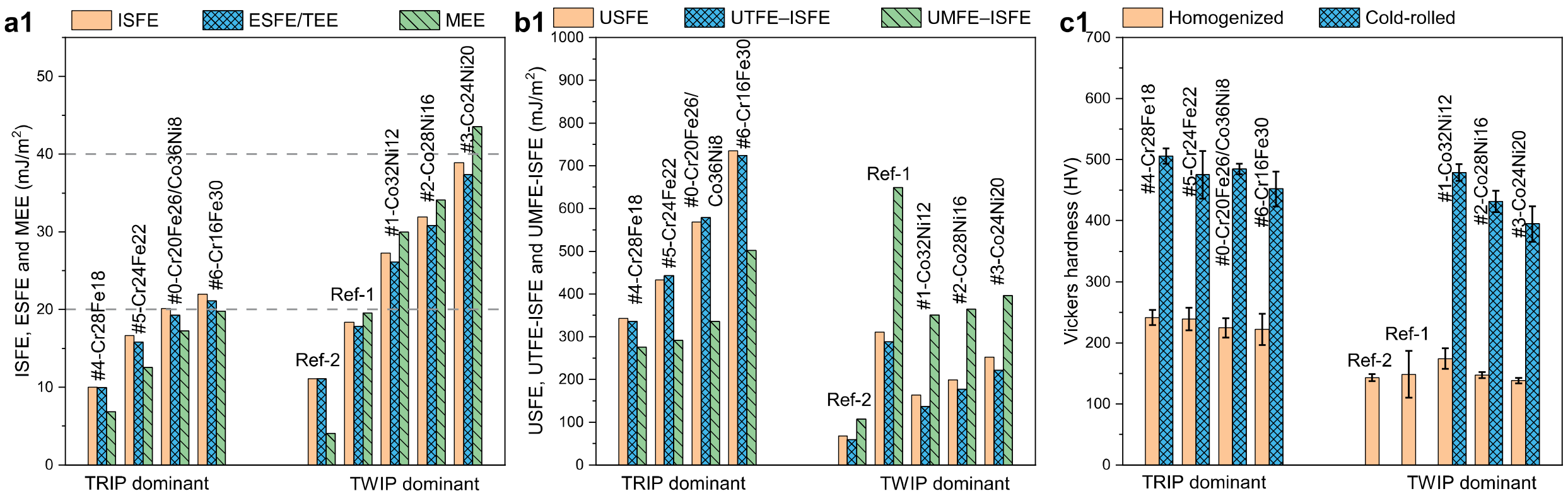

**a2**

| Nine alloys | ISFE range (mJ/m$^2$) | |
|---|---|---|
| Deformation type | < 20 | 20 - 40 |
| TRIP | 2 | 2 |
| TWIP | 2 | 3 |

**b2**

| Nine alloys | UMFE < UTFE | |
|---|---|---|
| Deformation type | True | False |
| TRIP | 4 | 0 |
| TWIP | 0 | 5 |

**c2**

| Nine alloys | High hardness | |
|---|---|---|
| Deformation type | True | False |
| TRIP | 4 | 0 |
| TWIP | 0 | 5 |

**Fig. 7 Details of fault energies, deformation mechanism, and hardness. a** Intrinsic stacking fault energy (ISFE), extrinsic stacking fault energy/twin embryo energy (ESFE/TEE), martensite embryo energy (MEE) of all designed alloys, and two reference alloys, the dashed lines are 20 and 40 mJ/m$^2$ corresponding to the upper limit of TRIP and TWIP respectively. **b** Unstable stacking fault energy (USFE), unstable twin fault energy (UTFE), unstable martensite fault energy (UMFE) of all designed alloys, and two reference alloys. **c** The hardness of homogenized and cold-rolled samples. The average and sample standard deviation for designed alloys hardness is calculated based on ten indents for each sample. The hardness and standard deviation of Ref-1 ($Co_{20}Cr_{20}Fe_{20}Mn_{20}Ni_{20}$) are taken from literature[65,78–83]. The hardness and standard deviation of Ref-2 ($Co_{10}Cr_{10}Fe_{40}Mn_{40}$) are adopted from[84]. **a1-c1** Plots and **a2-c2** Contingency tables summarize the relationships between deformation mechanisms (TRIP/TWIP) with ISF range, unstable fault energies, and hardness before and after cold rolling, respectively.

To examine if the new criterion can be more generally used for other types of alloys, DFT calculations of the different fault energies were performed for literature reported TRIP or TWIP steels[86–89] (Supplementary Table 11). The ISFE of TWIP steel[86] is calculated to be -289.2 mJ/m$^2$, and hence, we cannot predict the deformation mechanism based on the calculated ISFE. However, its UTFE (107.3 mJ/m$^2$) is much smaller than the UMFE (243.6 mJ/m$^2$), which indicates that the energy barrier for twinning is lower than martensitic transformation. In the high Mn steel $Mn_{23.7}Al_{5.2}Si_{5.5}Fe_{65.6}$, where both twinning and martensitic transformation operate during deformation[87], the UTFE and UMFE are 317.4 and 329.6 mJ/m$^2$, respectively. The similar energy barriers explain the coexistence of TRIP and TWIP well. However, it is difficult to infer the deformation mechanisms based on the ISFE, which is 27.8 mJ/m$^2$. In the four different steels tested in this work, the ISFE criterion cannot explain the deformation mechanisms, while all experimental observations can be explained by using the twinning and martensitic transformation energy barriers proposed in this work.

In conclusion, we find that processing history and athermal phase transformations should be considered in addition to chemical composition for an accurate phase prediction for HEAs. While a low ISFE is favorable for secondary deformation mechanism to operate, it cannot predict if martensitic transformation or twinning is the dominant deformation mechanism. Instead, the difference between UMFE/UTFE and ISFE is the energy barrier that determines the competition between twinning and martensitic transformation. The design route that combines rapid CALPHAD screening of phase stability and accurate DFT-based prediction of deformation mechanism is proven to be successful in this work and can accelerate the discovery of TRIP/TWIP in a wide spectrum of alloys.

## Data availability

The data that support the findings of this study are available from the corresponding author upon reasonable request.

## Code availability

The code that supports the findings of this study is available from the corresponding author upon reasonable request.

## Acknowledgment

The authors are grateful for helpful discussions with Mr. Noah Sargent, Dr. Yunhao Zhao and Dr. Hanlei Zhang. This material is based upon work supported by the National Science Foundation under Grants No. CMMI-2047218 and No. DMR-1945380. This work used the Extreme Science and Engineering Discovery Environment (XSEDE), which is supported by National Science Foundation grant number ACI-1548562. This research used resources of the National Energy Research Scientific Computing Center (NERSC), a U.S. Department of Energy Office of Science User Facility located at Lawrence Berkeley National Laboratory, operated under Contract No. DE-AC02-05CH11231.

## Author contributions

X. Wang proposed the alloy screening strategy, performed CALPHAD calculation, and designed the alloy composition (supervised by W. Xiong). R.T. Rodríguez De Vecchis prepared the sample, performed mechanical testing and microstructure characterization together with S. Sridar (supervised by W. Xiong) and X.B. Hu also helped in the experiments analysis. C. Li conducted the DFT calculation (supervised by W. Chen). All authors discussed the results and contributed to the writing of the paper.

## Competing interests

The authors declare no competing financial or non-financial interests.

# Supplementary materials

## Supplementary tables:

**Supplementary Table 1.** Details of the accuracy of different methods in phase prediction, which corresponds to Fig. 1a [1–4].

| Methods | | Classification types | Number of samples | Accuracy | References |
|---|---|---|---|---|---|
| Machine learning | Artificial neural network (ANN) | SS (solid solution), IM (intermetallic), SS+IM | 402 | 74 %* | 1 |
| | Support vector machine with a radial basis function | SS, NSS (Nonsolid solution) | 550 | 89 %** | 2 |
| | Neural network | bcc, fcc, fcc + bcc from SS | 550 | 91 %** | 2 |
| | Gaussian process classification | fcc, bcc, hcp & multiple phases | 1252 | 93 %** | 3 |
| CALPHAD | TCNI8 database | fcc, bcc, hcp & multiple phases | 77 | 94 % | 3 |
| | TCHEA1 database | fcc, bcc, hcp & multiple phases | 216 | 71 % | 4 |
| Phenomenological | Phenomenological parameter $\gamma$ | fcc, bcc, hcp & multiple phases | 124 | 73 % | 3 |

*The accuracy is evaluated via the 4-fold cross-validation
**The accuracy is evaluated via the 10-fold cross-validation

**Supplementary Table 2.** Details of the experiment measured intrinsic stacking fault energy in Fig. 1b, the mean value is taken for the same alloy reported in different publications [5–12].

| Alloy composition (Atomic) | Stacking fault energy ($mJ/m^2$) | Reference |
|---|---|---|
| $FeCoCrNiMo_{0.23}$ | 19 | B. Cai et al. [5] |
| NiCoCr | 22 ± 4 | G. Laplanche et al. [6] |
| $Fe_{20}Cr_{20}Mn_{20}Co_{27}Ni_{13}$ | 19 ± 3 | S.F. Liu et al. [7] |
| $Fe_{20}Cr_{20}Mn_{20}Co_{23}Ni_{17}$ | 25 ± 3 | S.F. Liu et al. [7] |
| CoCrFeMnNi | 26 ± 4 | S.F. Liu et al. [7] |
| CoCrFeMnNi | 30 | N. Okamoto et al. [8] |
| CoCrFeMnNi | 26.5 ± 4.5 | S. F. Liu et al. [9] |
| FeCoNiCr | 27 ± 4 | S. F. Liu et al. [9] |
| $(FeCoNiCr)_{86}Mn_{14}$ | 29 ± 4 | S. F. Liu et al. [9] |
| $(FeCoNiCr)_{94}Mn_{6}$ | 28 ± 4 | S. F. Liu et al. [9] |
| NiCoCr | 18 ± 4 | S. F. Liu et al. [9] |
| $Fe_{20}Co_{15}Ni_{25}Cr_{20}Mn_{20}$ | 38 ± 6 | S. F. Liu et al. [9] |
| $Fe_{40.4}Ni_{11.3}Mn_{34.8}Al_{7.5}Cr_{6}$ | 22.2 ± 1.9 | Z. W. Wang et al. [10] |
| $(Fe_{40.4}Ni_{11.3}Mn_{34.8}Al_{7.5}Cr_{6})_{99.3}C_{0.7}$ | 20.1 ± 1 | Z. W. Wang et al. [10] |
| $(Fe_{40.4}Ni_{11.3}Mn_{34.8}Al_{7.5}Cr_{6})_{98.9}C_{1.1}$ | 10.2 ± 0.9 | Z. W. Wang et al. [10] |
| Al0.1CoCrFeNi | 13.5 ± 7.5 | X.D. Xu et al. [11] |
| FeCoNiCr | 32.5 | Y. Wang et al. [12] |

**Supplementary Table 3.** Designed alloy composition, weighted alloy composition, and EDS measured alloy composition of sample $Co_{36}Cr_{20}Fe_{26}Mn_{10}Ni_{8}$.

| Element | Weight % | Target Mass (g) | Actual Mass (g) | Final W% | Δ(Wt%) % |
|---|---|---|---|---|---|
| Fe | 0.2578 | 10.3120 | 10.3122 | 0.2573 | -0.20 |
| Co | 0.3767 | 15.0680 | 15.0686 | 0.3759 | -0.20 |
| Ni | 0.0834 | 3.3360 | 3.3385 | 0.0833 | -0.13 |
| Cr | 0.1846 | 7.3840 | 7.3836 | 0.1842 | -0.21 |
| Mn | 0.0975 | 3.9000 | 3.9789 | 0.0993 | 1.81 |
| Total | 1 | 40.0000 | 40.0818 | 1.0000 | |

**Supplementary Table 4.** Designed alloy composition, weighted alloy composition, and EDS measured alloy composition of sample $Co_{32}Cr_{20}Fe_{26}Mn_{10}Ni_{12}$.

| Element | Weight % | Target Mass (g) | Actual Mass (g) | Final W% | Δ(Wt%) % |
|---|---|---|---|---|---|
| Fe | 0.2578 | 10.3130 | 10.3137 | 0.2573 | -0.18 |
| Co | 0.3349 | 13.3973 | 13.3922 | 0.3341 | -0.23 |
| Ni | 0.1251 | 5.0045 | 5.0068 | 0.1249 | -0.14 |
| Cr | 0.1847 | 7.3887 | 7.3886 | 0.1843 | -0.19 |
| Mn | 0.0976 | 3.9044 | 3.9789 | 0.0993 | 1.71 |
| Total | 1.0001 | 40.0080 | 40.0802 | 1.0000 | |

**Supplementary Table 5.** Designed alloy composition, weighted alloy composition, and EDS measured alloy composition of sample $Co_{28}Cr_{20}Fe_{26}Mn_{10}Ni_{16}$.

| Element | Weight % | Target Mass (g) | Actual Mass (g) | Final W% | Δ(Wt%) % |
|---|---|---|---|---|---|
| Fe | 0.2579 | 10.3160 | 10.3167 | 0.2574 | -0.20 |
| Co | 0.2931 | 11.7240 | 11.7279 | 0.2926 | -0.18 |
| Ni | 0.1668 | 6.6720 | 6.6768 | 0.1666 | -0.14 |
| Cr | 0.1847 | 7.3880 | 7.3846 | 0.1842 | -0.26 |
| Mn | 0.0976 | 3.9040 | 3.9785 | 0.0993 | 1.69 |
| Total | 1.0001 | 40.0040 | 40.0845 | 1.0000 | |

**Supplementary Table 6.** Designed alloy composition, weighted alloy composition, and EDS measured alloy composition of sample $Co_{24}Cr_{20}Fe_{26}Mn_{10}Ni_{20}$.

| Element | Weight % | Target Mass (g) | Actual Mass (g) | Final W% | Δ(Wt%) % |
|---|---|---|---|---|---|
| Fe | 0.2579 | 10.3160 | 10.3183 | 0.2572 | -0.26 |
| Co | 0.2512 | 10.0480 | 10.0847 | 0.2514 | 0.08 |
| Ni | 0.2085 | 8.3400 | 8.3402 | 0.2079 | -0.28 |
| Cr | 0.1847 | 7.3880 | 7.389 | 0.1842 | -0.27 |
| Mn | 0.0976 | 3.9040 | 3.9799 | 0.0992 | 1.66 |
| Total | 0.9999 | 39.9960 | 40.1121 | 1.0000 | |

**Supplementary Table 7.** Designed alloy composition, weighted alloy composition, and EDS measured alloy composition of sample $Co_{36}Cr_{28}Fe_{18}Mn_{10}Ni_{8}$.

| Element | Weight % | Target Mass (g) | Actual Mass (g) | Final W% | Δ(Wt%) % |
|---|---|---|---|---|---|
| Fe | 0.1794 | 1.7940 | 1.7962 | 0.1791 | -0.15 |
| Co | 0.3787 | 3.7870 | 3.7919 | 0.3781 | -0.15 |
| Ni | 0.0838 | 0.8380 | 0.8389 | 0.0837 | -0.17 |
| Cr | 0.2599 | 2.5990 | 2.6002 | 0.2593 | -0.23 |
| Mn | 0.0981 | 0.9810 | 1.0006 | 0.0998 | 1.72 |
| Total | 0.9999 | 9.9990 | 10.0278 | 1.0000 | |

**Supplementary Table 8.** Designed alloy composition, weighted alloy composition, and EDS measured alloy composition of sample $Co_{36}Cr_{24}Fe_{22}Mn_{10}Ni_{8}$.

| Element | Weight % | Target Mass (g) | Actual Mass (g) | Final W% | Δ(Wt%) % |
|---|---|---|---|---|---|
| Fe | 0.2187 | 2.1870 | 2.1881 | 0.2184 | -0.12 |
| Co | 0.3777 | 3.7770 | 3.7732 | 0.3767 | -0.27 |
| Ni | 0.0836 | 0.8360 | 0.8357 | 0.0834 | -0.21 |
| Cr | 0.2222 | 2.2220 | 2.2212 | 0.2217 | -0.21 |
| Mn | 0.0978 | 0.9780 | 0.9988 | 0.0997 | 1.95 |
| Total | 1 | 10.0000 | 10.0170 | 1.0000 | |

**Supplementary Table 9.** Designed alloy composition, weighted alloy composition, and EDS measured alloy composition of Sample $Co_{36}Cr_{16}Fe_{30}Mn_{10}Ni_8$.

| Element | Weight % | Target Mass (g) | Actual Mass (g) | Final W% | Δ(Wt%) % |
|---|---|---|---|---|---|
| Fe | 0.2966 | 2.9660 | 2.966 | 0.2958 | -0.28 |
| Co | 0.3756 | 3.7560 | 3.759 | 0.3749 | -0.20 |
| Ni | 0.0831 | 0.8310 | 0.8324 | 0.0830 | -0.11 |
| Cr | 0.1473 | 1.4730 | 1.4739 | 0.1470 | -0.21 |
| Mn | 0.0973 | 0.9730 | 0.9963 | 0.0994 | 2.11 |
| Total | 0.9999 | 9.9990 | 10.0276 | 1.0000 | |

**Supplementary Table 10.** Phase and grain boundary information in Fig. 5.

| Sample ID | Homogenized | | | 60 % Cold rolled | | | |
|---|---|---|---|---|---|---|---|
| | fcc | hcp | bcc | fcc | hcp | bcc | fcc_Σ3 |
| #Co36-Ni8/ Cr20-Fe26 | 0.998 | 0.001 | 0.001 | 0.509 | 0.421 | 0.07 | 0.003 |
| #Co32-Ni12 | 0.999 | 0 | 0 | 0.632 | 0.255 | 0.111 | 0.056 |
| #Co28-Ni16 | 0.999 | 0.001 | 0.001 | 0.854 | 0.099 | 0.046 | 0.071 |
| #Co24-Ni20 | 0.998 | 0.001 | 0.001 | 0.776 | 0.144 | 0.080 | 0.027 |
| #Cr28-Fe18 | 0.967 | 0.030 | 0.002 | 0.2 | 0.689 | 0.11 | 0.003 |
| #Cr24-Fe22 | 0.855 | 0.141 | 0.004 | 0.263 | 0.675 | 0.061 | 0.003 |
| #Cr16-Fe30 | 0.854 | 0.138 | 0.007 | 0.596 | 0.315 | 0.089 | 0.006 |

**Supplementary Table 11.** DFT calculated fault energies for literature reported[13–16] TRIP/TWIP austenite steel*

| Initial composition (at.%) | Adjusted composition (at.%) | Secondary deformation type | USFE ($mJ/m^2$) | ISFE ($mJ/m^2$) | UTFE ($mJ/m^2$) | UMFE ($mJ/m^2$) | Reference |
|---|---|---|---|---|---|---|---|
| $Mn_{22}C_3Fe_{75}$ | $Mn_{23}Fe_{77}$ | TWIP | 86.9 | -289.2 | 107.3 | 243.6 | 13 |
| $Mn_{23.7}Al_{5.2}Si_{5.5}Fe_{65.6}$ | $Mn_{23}Al_5Si_5Fe_{67}$ | TRIP/TWIP | 307.8 | 27.8 | 317.4 | 329.6 | 14 |
| $Mn_{1.7}Cr_{19.3}Ni_{7.6}Mo_{0.2}Cu_{0.3}N_{0.2}Si_{0.6}C_{0.2}Fe_{69.9}$ | $Mn_2Cr_{20}Ni_8Fe_{70}$ | TRIP | 478.6 | 34.6 | 497.2 | 254.6 | 15 |
| $Mn_{0.6}Cr_{17.9}Ni_{11.3}Mo_{1.5}Si_{1.2}C_{0.1}Fe_{67.4}$ | $Cr_{18}Ni_{11}Mo_2Si_2Fe_{67}$ | TWIP dominant | 243.2 | 17.4 | 277.8 | 324.1 | 16 |

*In order to improve the calculation efficiency, the low concentration (less than 2%) elements will be ignored in follow calculations. The C elements as interstitial impurities cannot be considered.

## Supplementary figures:

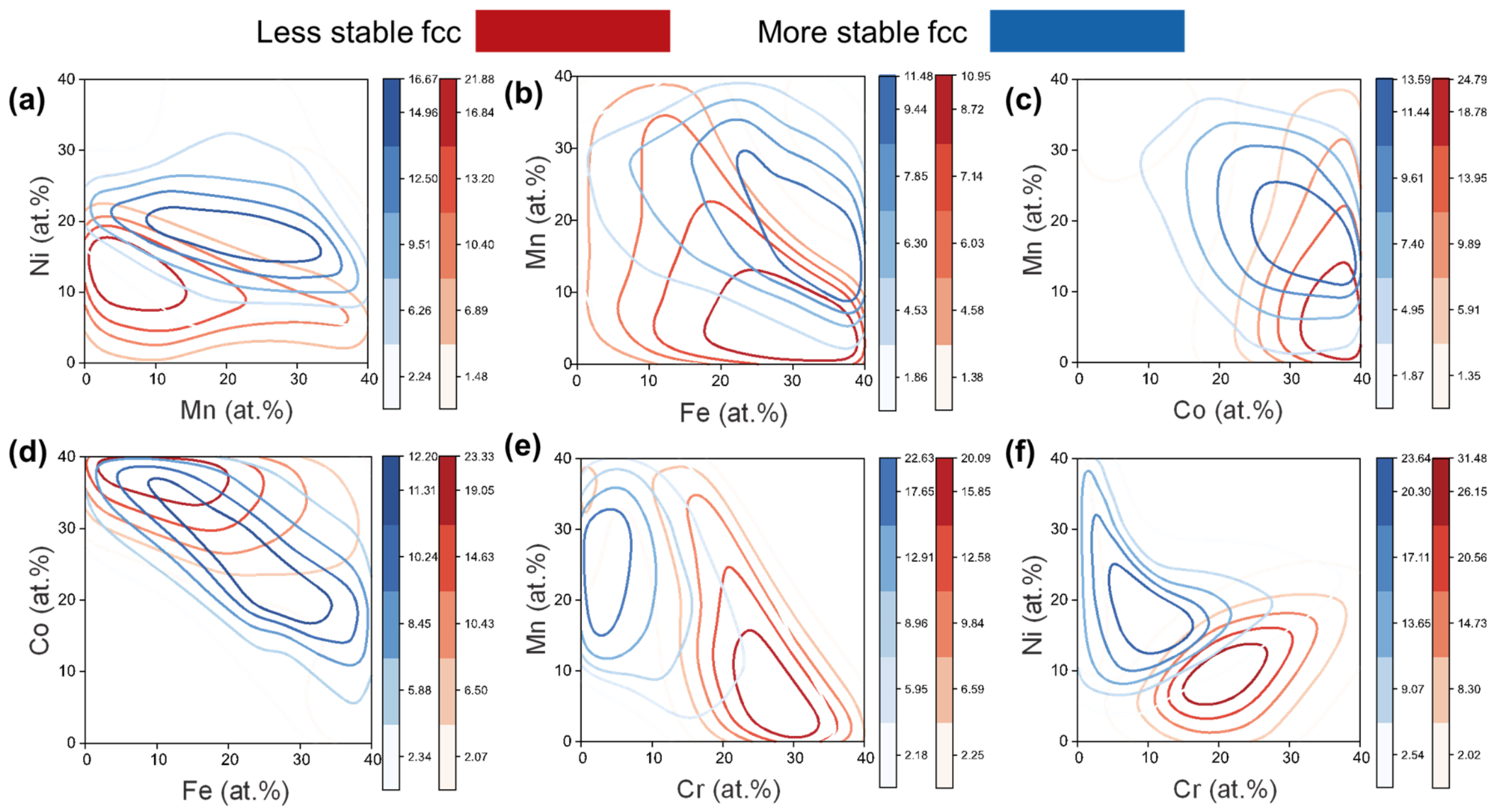

**Supplementary Figure 1.** The screened alloys have been classified as more unstable fcc alloy with red color($\Delta G_{m(fcc\rightarrow hcp)}^{HEAs} < \Delta G_{m(fcc\rightarrow hcp)}^{Co10Cr10Fe40Mn40}$), and less unstable fcc alloy with red color ($\Delta G_{m(fcc\rightarrow hcp)}^{Co10Cr10Fe40Mn40} < \Delta G_{m(fcc\rightarrow hcp)}^{HEAs} < \Delta G_{m(fcc\rightarrow hcp)}^{CoCrFeMnNi}$). The kernel density distribution of the two categories of alloys has been plotted with two varying elements, and a darker contour line indicates a higher density of the alloys in the composition region. (a) Mn-Ni pair, (b) Fe-Mn pair, (c) Co-Mn pair, (d) Fe-Co pair, (e) Cr-Mn pair, and (f) Cr-Ni pair.

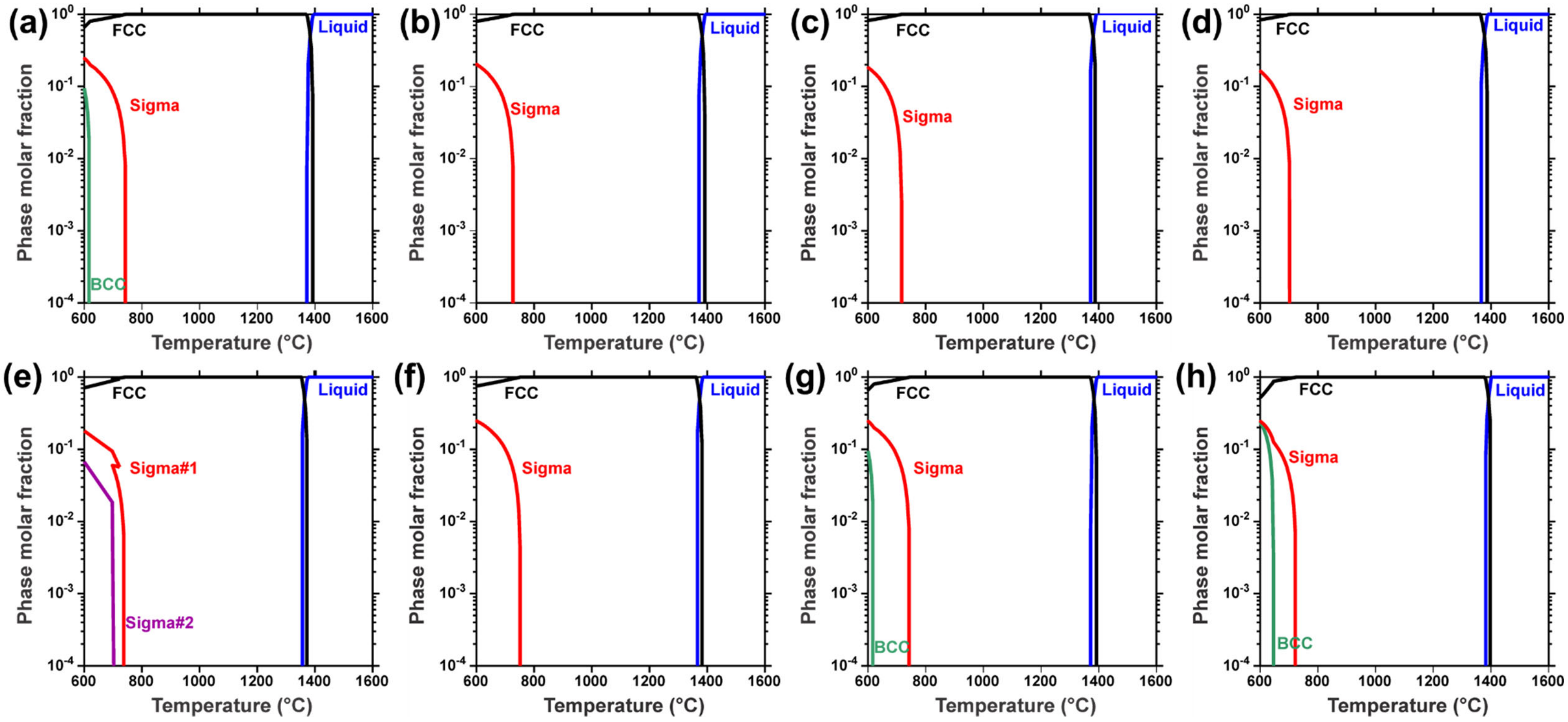

**Supplementary Figure 2.** Step diagrams of (a-d) alloys with the changing Co and Ni content, where (a) $Co_{36}Cr_{20}Fe_{26}Mn_{10}Ni_{8}$ (b) $Co_{32}Cr_{20}Fe_{26}Mn_{10}Ni_{12}$ (c) $Co_{28}Cr_{20}Fe_{26}Mn_{10}Ni_{16}$ (d) $Co_{24}Cr_{20}Fe_{26}Mn_{10}Ni_{20}$ (e-h) alloys with the changing Cr and Ni content, where (e) $Co_{36}Cr_{28}Fe_{18}Mn_{10}Ni_{8}$ (f) $Co_{36}Cr_{24}Fe_{22}Mn_{10}Ni_{8}$ (g) $Co_{36}Cr_{20}Fe_{26}Mn_{10}Ni_{8}$ (h) $Co_{36}Cr_{16}Fe_{30}Mn_{10}Ni_{8}$, please be noted that (a) and (g) are same.

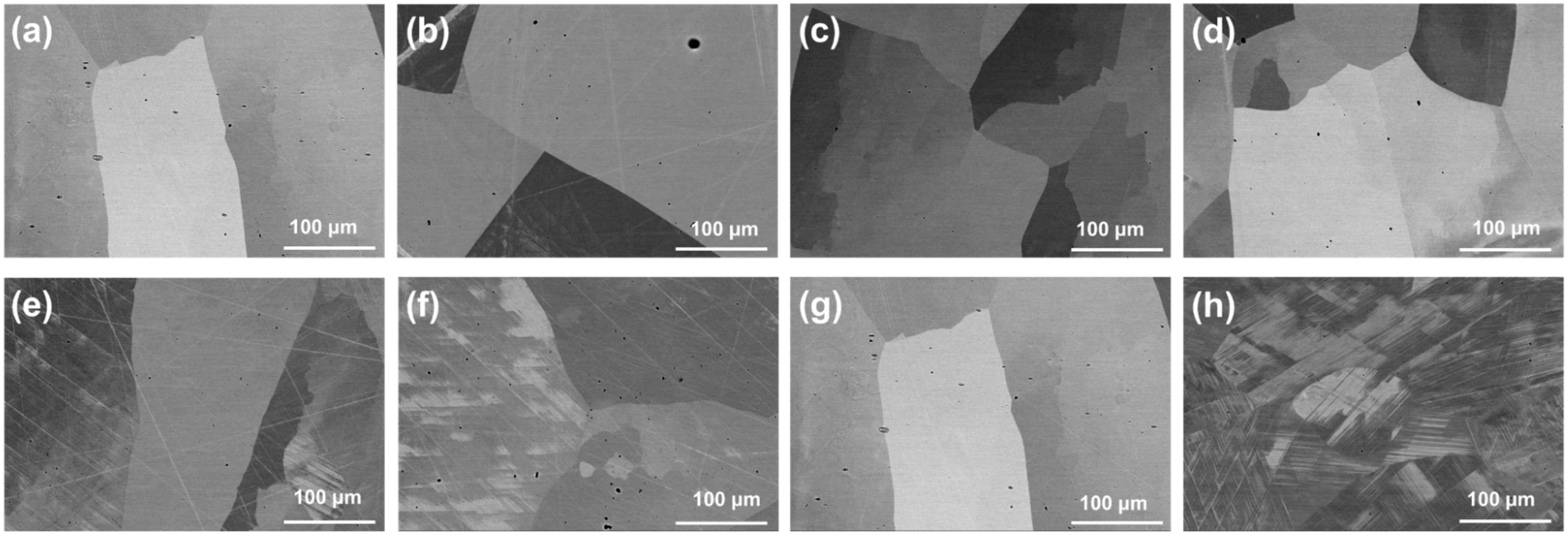

**Supplementary Figure 3.** SEM images of the homogenized sample (a)-(d) $Co_{36-x}Cr_{20}Fe_{26}Mn_{10}Ni_{8+x}$, where x = 0, 4, 8, 12. (e-f) $Co_{36}Cr_{28-x}Fe_{18+x}Mn_{10}Ni_{8}$, where x = 0, 4, 8, 12. (a) and (g) are the same sample.

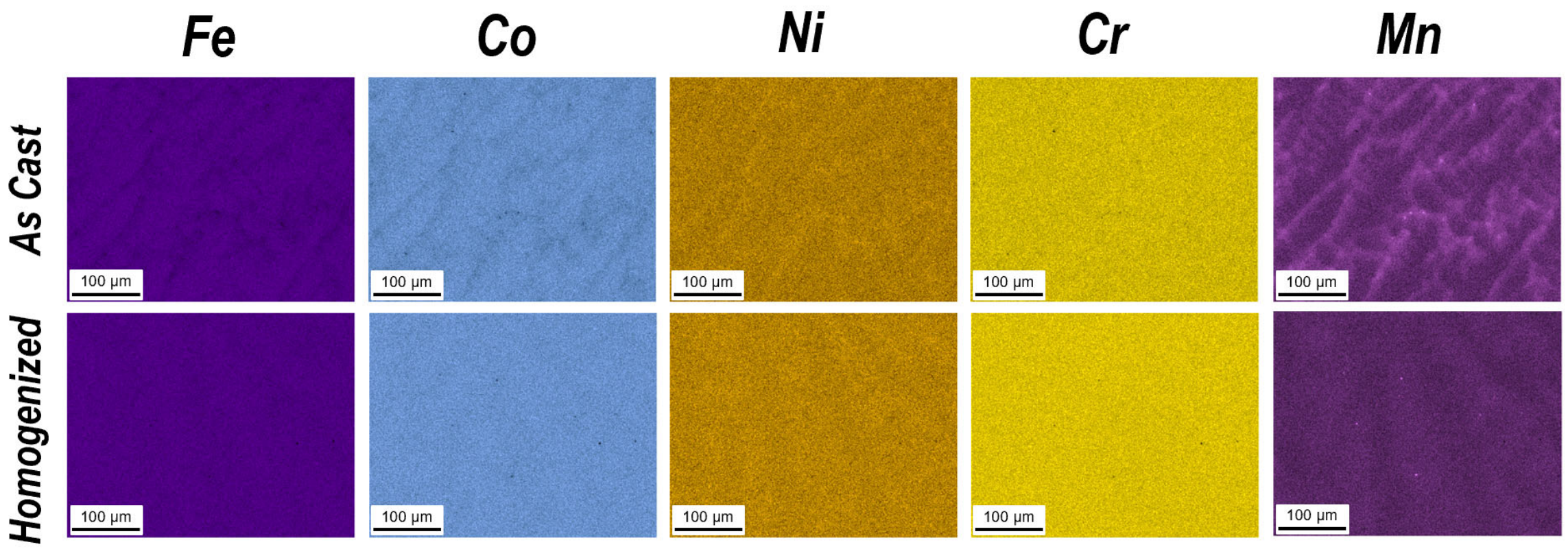

**Supplementary Figure 4.** EDS of as-cast and homogenized sample $Co_{36}Cr_{20}Fe_{26}Mn_{10}Ni_{8}$

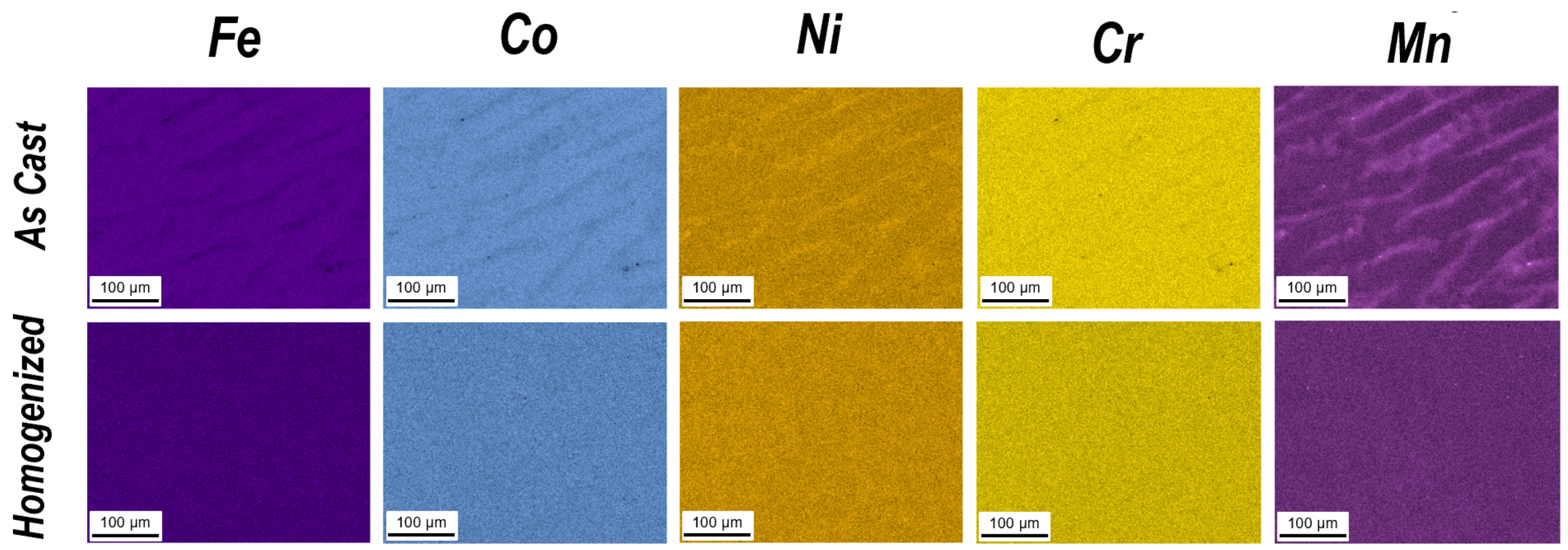

**Supplementary Figure 5.** EDS of as-cast and homogenized sample $Co_{32}Cr_{20}Fe_{26}Mn_{10}Ni_{12}$

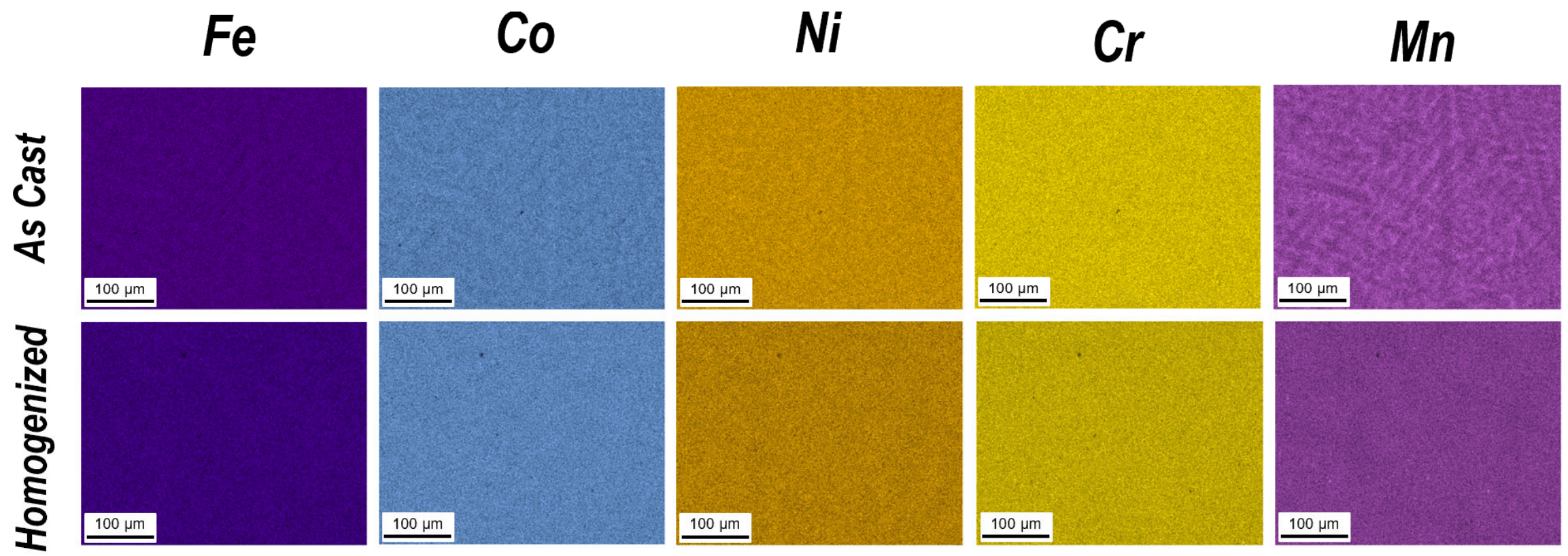

**Supplementary Figure 6.** EDS of as-cast and homogenized sample $Co_{28}Cr_{20}Fe_{26}Mn_{10}Ni_{16}$

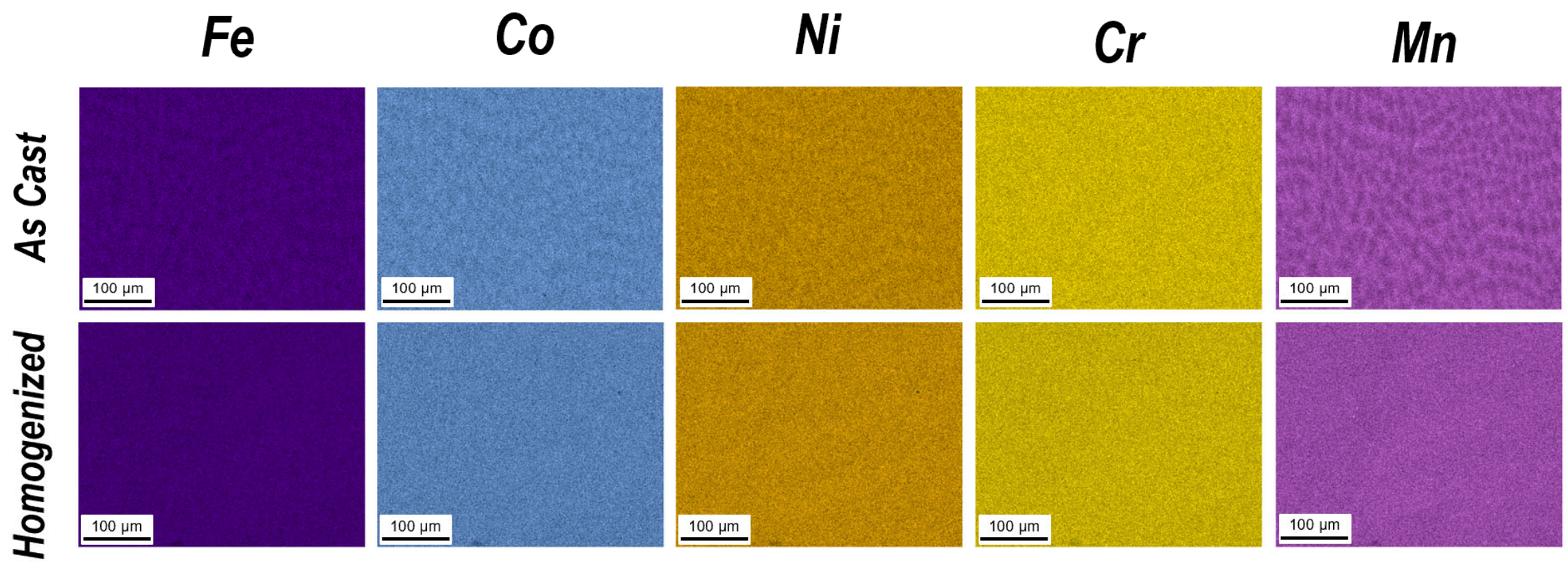

**Supplementary Figure 7.** EDS of as-cast and homogenized sample $Co_{24}Cr_{20}Fe_{26}Mn_{10}Ni_{20}$

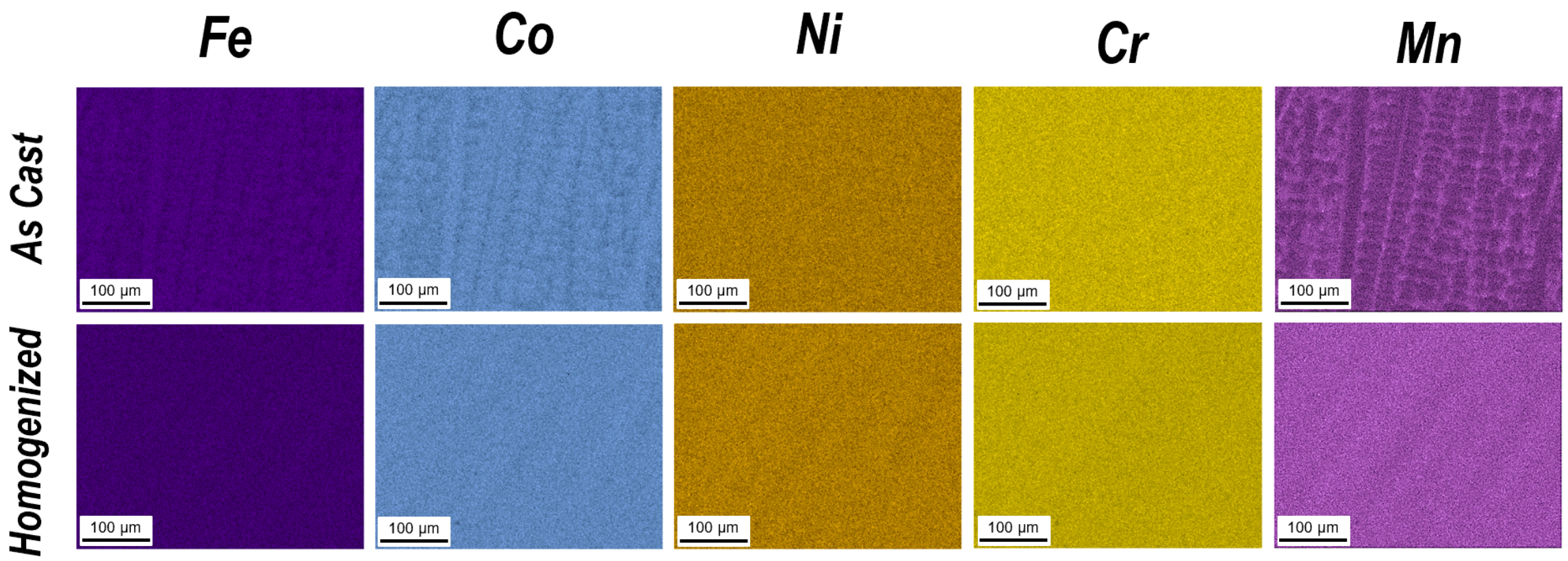

**Supplementary Figure 8.** EDS of as-cast and homogenized sample $Co_{36}Cr_{28}Fe_{18}Mn_{10}Ni_{8}$

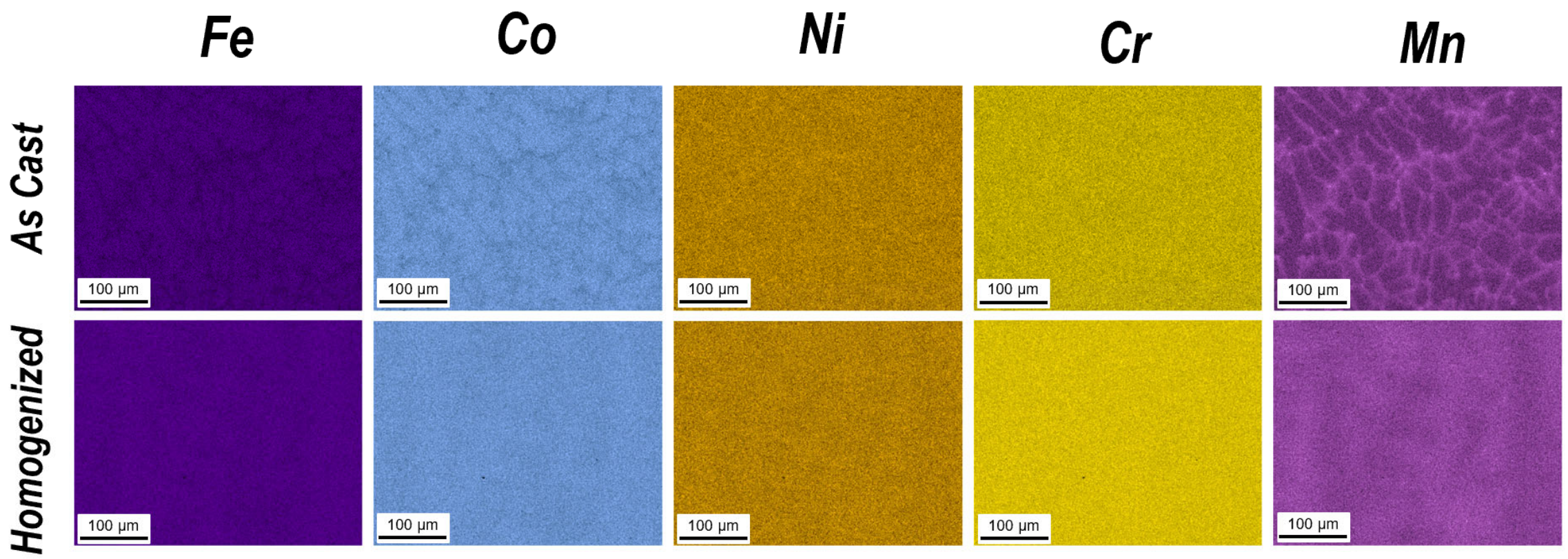

**Supplementary Figure 9.** EDS of as-cast and homogenized sample $Co_{36}Cr_{24}Fe_{22}Mn_{10}Ni_8$

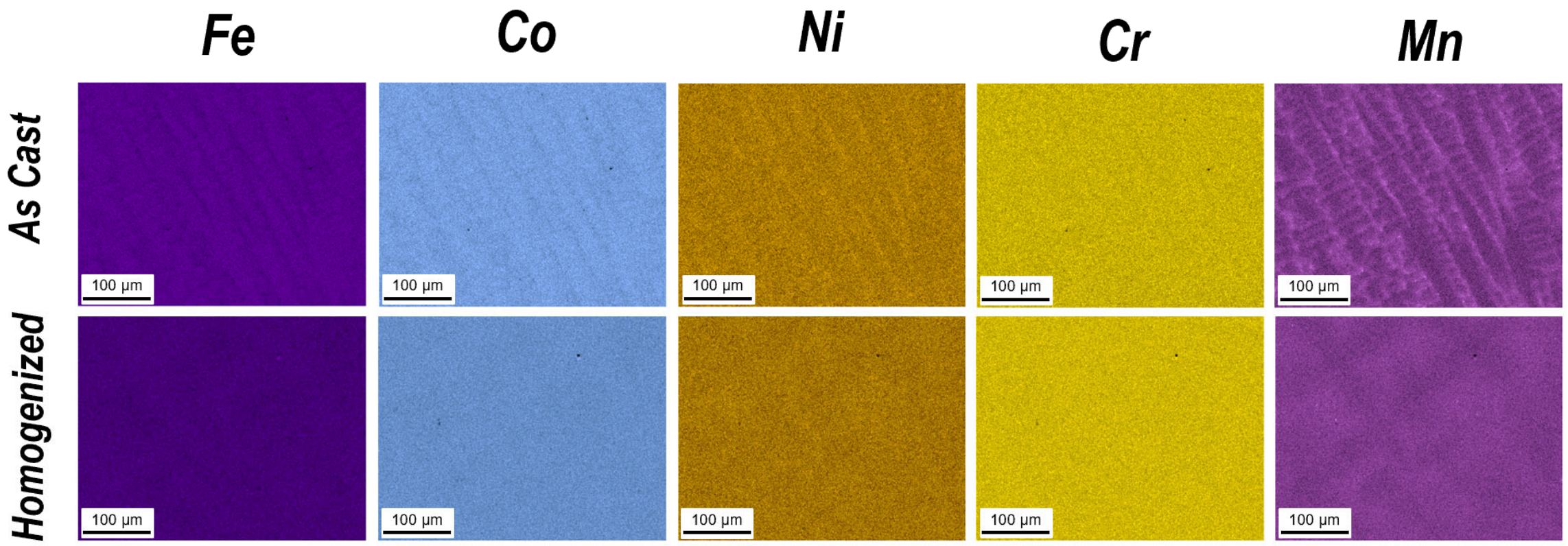

**Supplementary Figure 10.** EDS of as-cast and homogenized sample $Co_{36}Cr_{16}Fe_{30}Mn_{10}Ni_8$

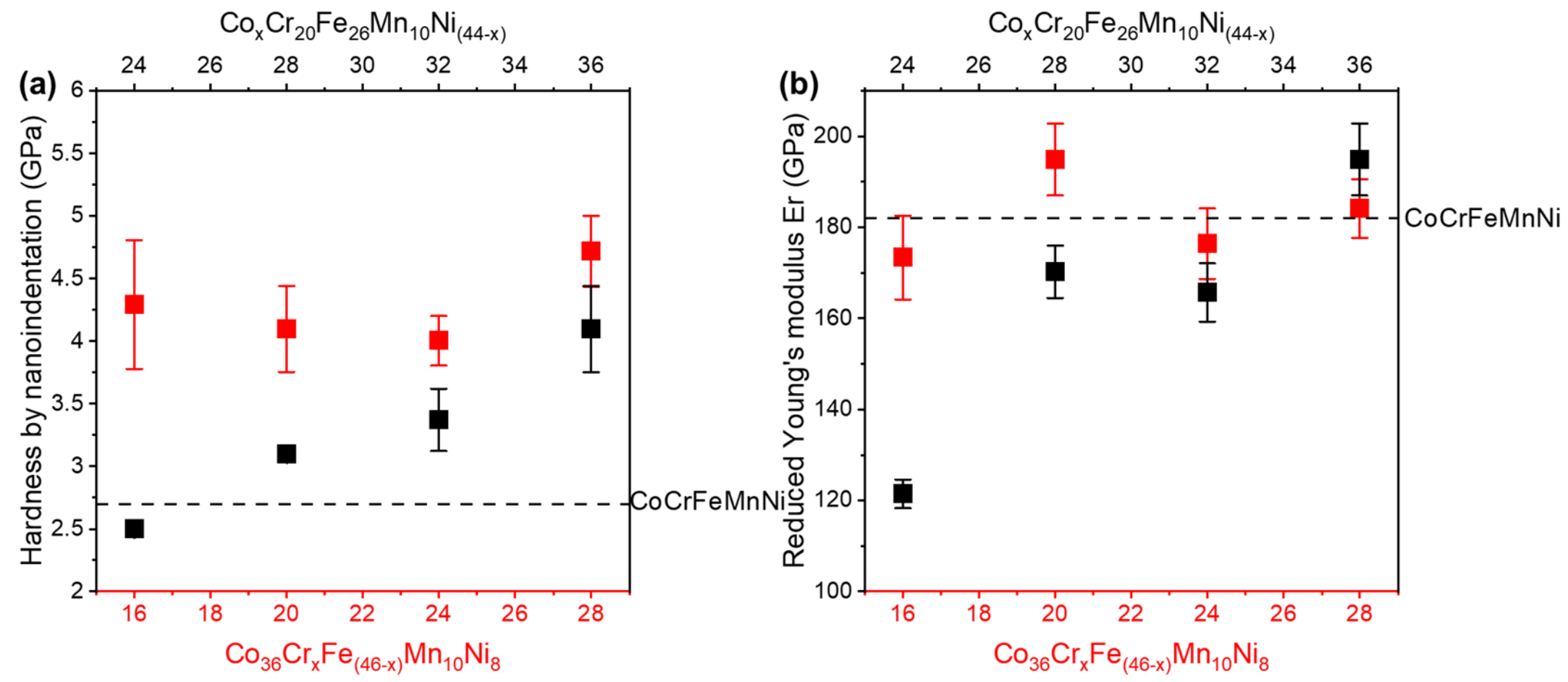

**Supplementary Figure 11.** Nanoindentation measured mechanical properties. (a) Hardness and (b) reduced Young's modulus.